\documentclass[journal,draftclsnofoot,onecolumn]{IEEEtran}
\usepackage{setspace}
\usepackage{amsmath}
\usepackage{amssymb}
\usepackage{mathrsfs}
\usepackage{amsthm}
\usepackage{multirow}
\usepackage{color}
\usepackage{longtable}
\usepackage{array}
\usepackage{url}
\usepackage{comment}
\usepackage{enumerate}
\usepackage{eucal}
\usepackage{graphics}
\usepackage{verbatim}
\usepackage{subfigure}

\usepackage{algorithm}
\usepackage{algpseudocode}
\usepackage{cite}
\usepackage[table]{xcolor}

\usepackage{stmaryrd}

\usepackage{mathtools}
\usepackage{xparse}
\DeclarePairedDelimiterX{\set}[1]{\{}{\}}{\setargs{#1}}
\NewDocumentCommand{\setargs}{>{\SplitArgument{1}{;}}m}
{\setargsaux#1}
\NewDocumentCommand{\setargsaux}{mm}
{\IfNoValueTF{#2}{#1} {#1\,\delimsize|\,\mathopen{}#2}}

\DeclarePairedDelimiter\abs{\lvert}{\rvert}

\DeclarePairedDelimiter\parenv{\lparen}{\rparen}

\DeclarePairedDelimiter\angenv{\langle}{\rangle}

\newtheorem{thm}{Theorem}[section]

\newtheorem{lem}{Lemma}[section]

\newtheorem{prop}{Proposition}[section]
\newtheorem{rmk}{Remark}[section]
\newtheorem{cor}{Corollary}[section]
\newtheorem{example}{Example}[section]

\newcommand{\RNum}[1]{\lowercase\expandafter{\romannumeral #1\relax}}
\newcommand{\Rnum}[1]{\uppercase\expandafter{\romannumeral #1\relax}}

\newcommand{\F}{\mathbb{F}}

\newcommand{\Q}{\mathbb{Q}}

\newcommand{\eqdef}{\triangleq}
\newcommand{\cA}{\mathcal{A}}
\newcommand{\cB}{\mathcal{B}}
\newcommand{\cC}{\mathcal{C}}
\newcommand{\cD}{\mathcal{D}}
\newcommand{\cE}{\mathcal{E}}

\newcommand{\cN}{\mathcal{N}}
\newcommand{\cO}{\mathcal{O}}

\newcommand{\cV}{\mathcal{V}}

\newcommand{\tF}{\widetilde{F}}
\newcommand{\tu}{\widetilde{u}}

\newcommand{\ty}{\widetilde{y}}

\newcommand{\tbf}{ \widetilde{\mathbf{f}}}
\newcommand{\tbg}{ \widetilde{\mathbf{g}}}

\newcommand{\tby}{ \widetilde{\mathbf{y}}}

\newcommand{\ba}{\mathbf{a}}

\newcommand{\bff}{\mathbf{f}}
\newcommand{\bg}{\mathbf{g}}

\newcommand{\br}{\mathbf{r}}
\newcommand{\bs}{\mathbf{s}}

\newcommand{\bu}{\mathbf{u}}

\newcommand{\bx}{\mathbf{x}}
\newcommand{\by}{\mathbf{y}}
\newcommand{\bz}{\mathbf{z}}
\newcommand{\bOne}{\mathbf{1}}
\newcommand{\Zero}{\mathbf{0}}
\newcommand{\Dataset}{\mathcal{D}}
\newcommand{\Asnset}{\mathcal{Z}}

\newcommand{\grad}{\mathbf{g}}

\newcommand{\bbta}{\pmb{\beta}}
\newcommand{\bmu}{\pmb{\mu}}

\newcommand{\In}{{\rm In}}
\newcommand{\Out}{{\rm Out}}
\newcommand{\tail}{{\rm tail}}
\newcommand{\head}{{\rm head}}
\newcommand{\wt}{{\rm wt}}

\newcommand{\Rank}{{\rm Rank}}
\newcommand{\mincut}{{\rm min\textup{-}cut }}

\newcommand{\sbinom}[2]{\genfrac{[}{]}{0pt}{}{#1}{#2}}

\begin{document}

\title{Linear Network Coding for Robust Function Computation and Its Applications in Distributed Computing}

\author{
  Hengjia Wei, Min Xu and  Gennian Ge%
  \thanks{This project was supported by   the National Key Research and Development Program of China under Grant 2020YFA0712100,
the National Natural Science Foundation of China under Grant 12231014 and Grant 12371523, Beijing Scholars Program, and 
the Pengcheng National Laboratory project under
Grant PCL2024AS103.}%
  \thanks{H. Wei (e-mail: hjwei05@gmail.com) is with the Peng Cheng Laboratory, Shenzhen 518055, China. He is also with the School of Mathematics and Statistics,
Xi’an Jiaotong University, Xi’an 710049, China, and the Pazhou Laboratory (Huangpu), Guangzhou 510555, China.}%
  \thanks{M. Xu (e-mail: minxu0716@qq.com) is with the School of Statistics and Data Science, LPMC \& KLMDASR, Nankai
University, Tianjin 300071, China.}%
\thanks{G. Ge (e-mail: gnge@zju.edu.cn) is with the School of Mathematical Sciences, Capital Normal University, Beijing 100048, China.}

}

\maketitle
\begin{abstract}
  We investigate linear network coding in the context of robust function computation, where a sink node is tasked with computing a target function of  messages generated at multiple source nodes. In a previous work, a new distance measure was introduced to evaluate the error tolerance of a linear network code for function computation, along with a Singleton-like bound for this distance. In this paper, we first present a minimum distance decoder for these linear network codes. We then focus on the sum function and the identity function, showing that in any directed acyclic network there are two classes of linear network codes for these target functions, respectively, that attain the Singleton-like bound. Additionally, we explore the application of these codes in distributed computing and design a distributed gradient coding scheme in a heterogeneous setting, optimizing the trade-off between straggler tolerance, computation cost, and communication cost. This scheme can also defend against Byzantine attacks. 
\end{abstract}

\begin{IEEEkeywords}
    Linear network coding, network function computation, sum networks, error correction, gradient coding
\end{IEEEkeywords}

\section{Introduction}
Network coding allows nodes within a network to encode the messages they receive and then transmit the processed outputs to downstream nodes. In contrast to simple message routing, network coding has the potential to achieve a higher information rate, which has garnered significant attention over the past two decades. When the encoding function at each network node is linear, the scheme is referred to as \emph{linear network coding}. Li et al. \cite{LiYeuCai03} investigated the multicast problem, where a source node aims to send messages to multiple sink nodes, and showed that a linear network coding approach using a finite alphabet is sufficient to achieve the maximum information rate.  Koetter and M\'edard \cite{KoeMed03} introduced an algebraic framework for linear network coding. Jaggi et al. \cite{JagSanChoEffEgnJaiTol05}  demonstrated that there is a polynomial-time algorithm for constructing maximum-rate linear network codes, provided that the field size is at least as large as the number of sink nodes.

Network communications can encounter various types of errors, such as random errors from channel noise, erasure errors due to traffic congestion, and malicious attacks by adversaries. Error correction in network communications is more complex than in traditional point-to-point communications, as even a single error on one link can spread to all downstream links, potentially corrupting all messages received by a sink node. Cai and Yeung \cite{2002CaiYeung,2006YCpart1,2006YCpart2} addressed this issue by integrating network coding with error correction  and introduced a new coding technique called \emph{network error-correction coding}, which mitigates errors by adding redundancy in the spatial domain rather than the temporal domain. In \cite{2006YCpart2,2011YYN,2006YCpart1}, three well-known bounds from classical coding theory, including the Singleton bound, are extended to network error-correction coding. Various methods have been proposed in \cite{2008Zhang,2011YYN,Mat11,GuangFuZhang2013,GuaYeu21} to construct linear network codes that meet the Singleton bound.

This work focuses on linear network coding for robust function computation. In this scenario, a sink node is required to compute a target function of source messages which are generated at multiple source nodes, while accounting for the possibility that communication links may be corrupted by errors. Each intermediate node can encode the messages it receives and transmit the resulting data to downstream nodes. Multiple communication links between any two nodes are permitted, under the reasonable assumption that each link has a limited (unit) communication capacity. The \emph{computing rate} of a network code is defined as the average number of times the target function can be computed without error per use of the network. The maximum achievable computing rate is known as  \emph{robust computing capacity}, or \emph{computing rate} in the error-free case. Some upper bounds on computing capacity  were provided in \cite{2011AFKZ,2013AFKZ,2018HTYG,2019GYYL} for the error-free case, with achievability demonstrated for certain network topologies and target functions.  These were recently extended in  \cite{WeiXuGe23} to account for robust computing capacity.

For general network topologies and target functions, characterizing the (robust) computing capacity is a challenging problem. In this paper, we focus on linear target functions, specifically $f(\bx)=\bx \cdot T$, where $T \in \F_q^{s\times l}$.  In the error-free model,  it has been proved that  linear network coding can achieve the computing capacity for an arbitrary directed acyclic network when  $l\in \set{1,s}$, see \cite{2011AFKZ,2012RD}. However, for $2\leq l \leq s-1$, determining the computing rate for a generic network remains an open problem.  For robust computing, the authors of this paper proposed a new distance measure in \cite{WeiXuGe23} to assess the error tolerance of a linear network code and derived a Singleton-like bound on this distance. In the same paper, we also demonstrated that this bound is tight when the target function is the sum of source messages and the network is a three-layer network. However,  it is still unclear whether this Singleton-like bound can be achieved in general networks.


In this paper, we continue the study on linear network coding for robust function computation and design linear network codes that meet the Singleton bounds. Additionally, we explore the applications of these codes in distributed computing, where a computation task is divided into smaller tasks and distributed across multiple worker nodes. Our contributions are as follows:
\begin{enumerate}
\item In Section~\ref{Sec:mindisdec}  we present a decoder for linear network codes designed for robust computing. While this decoder is based on the minimum distance principle and may involve high time complexity, it offers valuable insights into the workings of robust network function computation.

\item In Section~\ref{Sec:sum_network} and Section~\ref{Sec:genfun}, we consider the sum function and the identity function, respectively. For these two target functions, we demonstrate the existence of linear network codes in any directed acyclic network with distances meeting the Singleton bound, assuming the field size is sufficiently large. Using these codes, in Section~\ref{Sec:capacity} we derive some lower bounds on the robust computing capacity for $f(\bx) =\bx \cdot T$ with $T\in \F_q^{s \times l}$.  In particular, when $l=1$ or $s$, we show that (scalar) linear network coding can either achieve the cut-set bound on robust computing capacity or match its integral part, respectively.

\item Section~\ref{Sec:application} establishes a connection between robust computing in a three-layer network and a straggler problem in the context of distributed computing, where a straggler refers to a worker node that performs  significantly slower than other nodes. By applying linear network codes for the sum function, we design a distributed gradient coding scheme in a heterogeneous setting, optimizing the trade-off between straggler tolerance, computation cost, and communication cost.
\end{enumerate}


\section{Preliminary}\label{sec:prelim}
\subsection{Network function computation model}
Let $G=(\mathcal{V},\mathcal{E})$ be a directed acyclic graph with a finite vertex set $\mathcal{V}$ and an edge set $\mathcal{E}$, where multiple edges are allowed between two vertices.
For any edge $e\in \mathcal{E}$, we use $\tail(e)$ and $\head(e)$ to denote the tail node and the head node of $e$.
For any vertex $v\in \mathcal{V}$, let $\In(v)=\set{e\in E;\head(e)=v}$ and $\Out(v)=\set{e\in E;\tail(e)=v}$, respectively.

In this paper, a \emph{network} $\cN$ over $G$ contains a set of \emph{source nodes} $S=\set{\sigma_1,\sigma_2,\ldots,\sigma_s}\subseteq \cV$ and a \emph{sink node} $\gamma \in \cV\setminus S$. Such a network is denoted by $\cN=(G,S,\gamma)$. Without loss of
generality, we assume that every source node has no incoming
edges. We further assume that there exists a directed path from every node $u \in \cV\setminus \set{\gamma}$ to $\gamma$ in $G$. Then it follows from the
acyclicity of $G$ that the sink node $\gamma$ has no outgoing edges.

In the network function computing problem, the sink node $\gamma$ needs to compute a \emph{target function} $f$ of the form \[f: \cA^s\longrightarrow\mathcal{O},\]
where $\cA$ and $\cO$ are finite alphabets, and the $i$-th argument of $f$
is generated at the  source node $\sigma_i$. 
Let $k$ and $n$ be two positive integers, and let $\cB$ be a finite alphabet. A $(k,n)$  \emph{network function computing code} (or network code for short) $\cC$ over $\cB$  enables the sink node $\gamma$ to compute the target function $f$ $k$ times by transmitting at most
$n$ symbols in $\cB$ on each edge in $\cE$, i.e., using the network
at most $n$ times.

In this paper, we focus on the problem of computing linear functions by linear codes. We assume that $\cA=\cB=\cO=\F_q$ and the target function has the form $f(\bx)=\bx\cdot T$ for some $s\times l$ matrix $T$ over $\F_q$, where $1\leq l\leq s$. Without loss of generality, we further assume that $T$ has full column rank, namely, its columns are linearly independent. Suppose that every source node $\sigma_i$ generates a  vector $\bx_i=(x_{i1},\cdots,x_{ik})$ of length $k$ over  $\cA$. Denote the vector of all the source messages by $\bx_S\eqdef (\bx_1,\cdots,\bx_s)$.
Computing the target function $k$ times implies that
the sink node requires
\begin{equation*}
  f(\bx_S)\eqdef \bx_S\cdot(T\otimes I_k),
\end{equation*}
where $I_k$ is the $k\times k$ identity matrix and $\otimes$ is the Kronecker product.

A $(k,n)$ network code is called \emph{linear} if the message transmitted by each edge $e$ is a linear combination of the messages received by $\tail(e)$.  In this paper, we mainly study the case of $n=1$, that is, the message transmitted  by each edge is an element of $\F_q$. Such a network code is known as a \emph{scalar} network code.
Specifically,  in a $(k,1)$ linear network code over $\F_q$, 
the message $u_e \in \F_q$ transmitted via edge $e$ has the form
\begin{equation}\label{eq:linlocalenc}
  u_e=
  \begin{cases}
    \sum\limits_{j=1}^{k} x_{ij} k_{(i,j),e}, & \mbox{if $\tail(e)= \sigma_i$ for some $i$}; \\
    \sum\limits_{d\in \In(\tail(e))} u_d k_{d,e}, & \mbox{otherwise},
  \end{cases}
\end{equation}
where  $k_{(i,j),e}, k_{d,e} \in \F_{q}$, and $k_{(i,j),e}$ is zero  if $e$ is not an outgoing edge of some source node $\sigma_i\in S$ and   $k_{d,e}$ is zero if $e$ is not an outgoing edge of $\head(d)$. 
 So, each $u_e$ can  be written as a linear combination of the source messages:
\begin{equation*}
  u_e = \bx_S \cdot \bff_e,
\end{equation*}
where $\bff_e \in \F_q^{sk}$.

Let $\by\in({\F_q})^{|\In(\gamma)|}$ be the message vector received by the sink node $\gamma$. Denote $F \eqdef (\bff_e:e\in \In(\gamma))$.
Then
\[
\by =  \bx_S \cdot F.
\]
The matrix $F$ is called the \emph{global encoding matrix}. Denote $K\eqdef (k_{d,e})_{d \in \cE, e\in \cE}$, and $B_i \eqdef ( k_{(i,j),e} )_{ j \in [k], e \in \cE}$ where $i=1,2,\ldots,s$ and $[k]$ denotes the set $\set{1,2,\ldots,k}$. In this paper, $K$ is referred to as \emph{transfer matrix} and $B_i$'s are referred to as \emph{source encoding matrices}.
For a subset of links $\rho \subseteq \cE$, let $A_\rho=(A_{d,e})_{d \in \rho, e\in \cE}$ where
\begin{equation*}
  A_{d,e}=
  \begin{cases}
    1, & \mbox{if $d=e$}; \\
    0, & \mbox{otherwise}.
  \end{cases}
\end{equation*}
Since the network is finite and acyclic, it is easy to see that the global encoding matrix
\begin{equation}\label{eq:matF}
F= \begin{pmatrix} B_1\\B_2\\ \vdots \\ B_s \end{pmatrix} (I-K)^{-1} A_{\In(\rho)}^\top.
\end{equation}

If there is a decoding function $\phi :\prod_{\In(\gamma)} \F_q^n \to \F_q^{kl}$ such that for all $\bx_S \in \F_q^{sk}$,
\begin{equation*}
    \phi\parenv*{\bx_S\cdot F} =\bx_S\cdot(T\otimes I_k),
\end{equation*}
then we say this (scalar) linear network code enables the sink node to compute the target function with rate $k$.

It is of particular interest to determine the maximum computing rate for a given network and a specific target function. An upper bound for this rate can be derived using the network's \emph{cut}. To proceed, we first introduce some necessary concepts. An edge sequence $(e_1,e_2,\cdots,e_n)$ is called a \emph{path} from node $u$ to node $v$ if $\tail(e_1)=u,\head(e_n)=v$ and $\tail(e_{i+1})=\head(e_i)$ for $i=1,2,\cdots,n-1$. For a vertex $v$ and a path $P$, we say $v\in P$ if there is an edge $e\in P$ such that $\tail(e)=v$ or $\head(e)=v$.
For two nodes $u,v\in \mathcal{V}$, a \emph{cut} of them is an edge set $C$ such that every path from $u$ to $v$ contains at least one edge in $C$.
 If $C$ is a cut of $\gamma$ and some source node $\sigma_i$, then we simply call it a \emph{cut of the network}.
Let $\Lambda(\mathcal{N})$ be the collection of all cuts of the network $\mathcal{N}$.
For a cut $C\in \Lambda(\cN)$, define
\[
    I_C   \eqdef \set{\sigma\in S ;  \mbox{there is no path from $\sigma$ to $\gamma$ after deleting the edges in $C$ from $\cE$}}.
\]
\begin{lem}[{\cite[Corollary~II.1]{WeiXuGe23}}]\label{lem:csbnd}
Given a network $\mathcal{N}$ and a linear target function $f(\bx)=\bx \cdot T$ with $T\in \F_q^{s\times l}$.  If there exists a $(k,n)$ network code $\cC$ computing $f$ with rate $k/n$, then  necessarily
\[ k/n  \leq  \min_{C\in \Lambda(\cN)} \frac{\abs{C}}{  {\rm{Rank}}(T_{I_C})},\]
where $T_{I_C}$ is the $\abs{I_C}\times l$ submatrix of $T$  which is obtained by choosing the rows of $T$ indexed by  $I_C$.
\end{lem}

For the sum function, i.e., $f(\bx)=\sum_{i=1}^s x_i$, Ramamoorthy demonstrated in  \cite[Theorem 2]{2008Ramamoorthy} that the upper bound presented in  Lemma~\ref{lem:csbnd} can be achieved using linear network coding. For the identity function, i.e., $T=I_s$, Rasala Lehman and Lehman showed in \cite[Theorem 4.2]{lehman2004}  that this bound can be attained  simply through  routing.
For a target function $f(\bx)=\bx\cdot T$ with $l=s-1$,  
Appuswamy and Franceschetti  \cite{2014Appuswamy} explored the achievability of a computing rate of one and showed that 
the   condition $1 \leq \min_{C\in \Lambda(\cN)} \frac{\abs{C}}{  {\rm{Rank}}(T_{I_C})}$  in Lemma~\ref{lem:csbnd}  is also sufficient. For general linear target functions with $2\leq l \leq s-2$, the achievability  of the bound in Lemma~\ref{lem:csbnd} remains an open problem.  

\subsection{Robust network function computation model}
\label{subsec:RNFCmodel}

Let $u_e\in \F_q$ be the message that is supposed to be transmitted by a link $e$.
If there is an error in  $e$, the message  transmitted by $e$, denoted by $\widetilde{u}_e$, can be written as $\widetilde{u}_e = u_e + z_e$ for some $z_e\in \F_q$.
We treat $z_e$ as a message, called \emph{error message}, and the vector $\bz=(z_e: e\in \cE)$ is referred to as an {error  vector}. An \emph{error pattern} $\rho$ is a set of links in which errors occur. We say an error  vector $\bz$ \emph{matches} an error pattern $\rho$, if $z_e=0$ for all $e \notin\rho$. 


According to \eqref{eq:linlocalenc}, $\tu_e$ has the following form.
\begin{equation*}
  \widetilde{u}_e=
  \begin{cases}
    \sum\limits_{j=1}^{w} x_{ij} k_{(i,j),e} + z_e, & \mbox{if $\tail(e)= \sigma_i$ for some $i$}; \\
    \sum\limits_{d\in \In(\tail(e))}\widetilde{u}_d k_{d,e} + z_e, & \mbox{otherwise}.
  \end{cases}
\end{equation*}
It can also be written as a linear combination of the source messages  and the errors, i.e.,
\begin{equation*}
  \widetilde{u}_e = (\bx_S, \bz)\cdot \tbf_e,
\end{equation*}
where $\tbf_e$ is known as \emph{extend global encoding vector.}
Let $\widetilde{\mathbf{y}}\in({\F_q})^{|\In(\gamma)|}$ be the message vector received by the sink node $\gamma$. Denote $\widetilde{F} \eqdef (\tbf_e:e\in \In(\gamma))$.
Then
\[  \widetilde{\mathbf{y}}=  (\bx_S, \bz )\cdot\widetilde{F},
\]
and $\tF$ is called the \emph{extended global encoding matrix}.
We may write
\begin{equation}\label{eq:extencmtx}
\widetilde{F} =\begin{pmatrix}
 F \\G
\end{pmatrix},\end{equation}
where $F$ is the global encoding matrix and $G$ is an $\abs{\cE}\times \abs{\In(\gamma)}$ matrix over $F_q$ which satisfies:
\begin{equation}\label{eq:matG}
G = (I-K)^{-1} A_{\In(\rho)}^\top.
\end{equation}

For a linear network code $\cC$ which can compute the function $f(\bx)=\bx \cdot T$ with rate $k$, we say it is \emph{robust to $\tau$ erroneous links} if
\[ \bx_S F +\bz G \neq \bx_S' F +\bz' G \]
for any $\bx_S,\bx_S'\in \F_q^{sk}$ and $\bz,\bz'\in \F_q^{\abs{\cE}}$ with $\bx_S (T\otimes I_k ) \neq \bx_S' (T\otimes I_k )$ and $\wt_H(\bz),\wt_H(\bz')\leq \tau$.

Denote
\[
  \Phi\eqdef \set{\bx\cdot F  ; \bx(T\otimes I_k)\neq\mathbf{0},\  \bx\in\F_q^{sk}}
\]
and
\[\Delta(\rho) \eqdef \set{\mathbf{z}\cdot G|\  \mathbf{z}\in\F_q^{|\mathcal{E}|} \textup{ matching the error pattern }\rho}.\]
Note that $\Zero \notin \Phi$. The \emph{minimum distance} of the network code $\mathcal{C}$ which computes the function $f(\bx)=\bx \cdot T$ with rate $k$ is defined as
\begin{equation}\label{eq:disdef}
    d_{\min}(\cC,T,k)\eqdef \min\set{\abs{\rho};\Phi\cap\Delta(\rho)\neq\varnothing}.
\end{equation}

The following result shows that the minimum distance defined above can be used to measure the error tolerance of $\cC$ for the target function $f(
\bx)=\bx \cdot T$.

\begin{thm}[{\cite[Theorem IV.1]{WeiXuGe23}}]\label{thm:minimum_distance}
Let $\tau$ be a positive integer. For a linear network code $\cC$ with target function $f(x)=\bx \cdot T$ and computing rate $k$, it  is robust to any error $\bz$ with $\wt_H(\bz)\leq \tau$ if and only if   $d_{\min}(\cC,T,k)\geq 2\tau+1$.
\end{thm}

In \cite{WeiXuGe23}, the authors derived a Singleton-like bound on $d_{\min}(\cC,T,k)$.

\begin{thm}[{\cite[Theorem IV.2]{WeiXuGe23}}]\label{thm:Sinbnd}
Given a network $\mathcal{N}$ and a target function $f(\bx)=\bx \cdot T$. Let $k$ be a positive integer. If there is a linear network code $\cC$ computing $f$ with rate $k$, then
\begin{equation*}
    d_{\min}(\cC,T,k)\leq\min\limits_{C\in\Lambda(\mathcal{N})}\set*{\abs{C}-k\cdot{\rm{Rank}}(T_{I_C})+1},
\end{equation*}
where $T_{I_C}$ is the $\abs{I_C}\times l$ submatrix of $T$  corresponding to the source nodes in  $I_C$.
\end{thm}

It has been shown in \cite[Theoerm~IV.4]{WeiXuGe23} that in a multi-edge tree network, for any linear target function $f(\bx) =\bx \cdot T$, this bound can be achieved if the field size is large enough.

In this paper, we focus on the cases where $l \in\set{1,s}$ and explore the achievability of the Singleton-like bound in arbitrary directed acyclic networks. For $l=1$, the target function can be expressed as $f(\bx)=\sum_{i} t_i x_i$. W.l.o.g., we assume that  each $t_i \neq 0$.  Then ${\rm{Rank}}(T_{I_C})=1$ for every cut $C \in \Lambda(\cN)$. We use $\mincut(u,v)$ to denote the size of the minimal cut between two nodes $u$ and $v$.  The Singleton-like bound then reads:
\[d_{\min}(\cC,T,k)\leq\min\limits_{\sigma_i\in S} \set*{\mincut(\sigma_i,\gamma) -k+1}. \]
Since computing the function $f(\bx)=\sum_{i} t_i x_i $ can be reduced to computing the sum by multiplying each source message $x_i$ by a scalar $t_i$, it suffices to consider the sum function, i.e.,  $T =\bOne$. 
 
In \cite{WeiXuGe23}, we studied a three-layer network with $s$ source nodes in the first layer, $N$ intermediate nodes in the second layer, and a single sink node in the third layer. Each source node is connected to some intermediate nodes in the second layer, and all intermediate nodes are connected to the sink node in the third layer. It is proven in \cite{WeiXuGe23} that for the sum function and any arbitrary three-layer network, the Singleton-like bound can be achieved as long as the field size is larger than the number of intermediate nodes.

\begin{thm}[{\cite[Theorem~IV.3]{WeiXuGe23}}]
\label{thm:3layernetwork}
Let $\cN$ be a three-layer network. Let $c^*=\min_{\sigma_i \in S} \abs{\Out(\sigma_i)}$ be the minimum out-degree of the source nodes\footnote{In a three-layer network, we have $\Out(\sigma_i) = \mincut(\sigma_i,\gamma)$.}. Assume that $q-1\geq N$. Then there is a linear network code  $\cC$ over $\F_q$ which can compute the sum of the source messages with rate $k$ and minimum distance
\[d_{\min}(\cC,\bOne, w) = c^*-k+1.\]
\end{thm}
In Section~\ref{Sec:sum_network}, we generalize this result and show that the Singleton-like bound can be achieved for the sum function in any directed acyclic network, provided that  the field size is sufficiently large. 

In Section~\ref{Sec:genfun}, we study the case of $l=s$.  Since $T\in \F_q^{s\times s}$ has full rank, we have $\Rank(T_{I_C}) = \abs{I_C}$ for any $C \in \Lambda(\cN)$. The Singleton-like bound then reads 
\[d_{\min}(\cC,T,k)\leq\min\limits_{C\in\Lambda(\mathcal{N})}\set*{\abs{C}-k\cdot \abs{{I_C}}+1}.\]
We will show that for the identity function, i.e., $T=I$, this bound is  achievable. Since the sink node can compute $\bx \cdot T$ as long as it recovers $\bx$, this conclusion also applies to any invertible matrix $T \in \F_q^{s\times s}$.


Given a network $\mathcal{N}$ with a target function $f$ and an error-tolerant capability $\tau$, the \emph{robust computing capacity} is defined as
\begin{align*}
C(\mathcal{N},f,\tau)\eqdef \sup \{k/n \mid  \textup{ there is a $(k,n)$ network code that} \textup{ can compute $f$ against $\tau$ errors} \}.
\end{align*}
For $\tau=0$, the robust computing capacity is also referred to as \emph{computing capacity}. Some upper bounds on robust computing capacity have been derived in \cite{WeiXuGe23}. 
We will use the linear network codes presented in Section~\ref{Sec:sum_network} and Section~\ref{Sec:genfun}, along with the time-sharing technique, to derive some lower bounds on the robust computing capacity for any linear target function $f(\bx)=\bx \cdot T$. Notably, when $l=1$, the lower bound meets the upper bound; when $l=s$, the lower bound can achieve the integral part of the upper bound.

\section{Decoding for Robust Network Function computation}\label{Sec:mindisdec}
In this section, we present a minimum distance decoder to illustrate the mechanism of robust network function computation. We first define a new metric.  Let $\cC$ be a linear network code for a network $\cN$ to compute a target function $f(\bx)=\bx\cdot T$. For two vectors $\by_1,\by_2 \in \F_q^{\abs{\In(\gamma)}}$,
their  \emph{distance with respect to $\cC$}, denoted by $d_{\cC}(\mathbf{y_1},\mathbf{y_2})$, is defined as
\begin{equation*}
d_{\cC}(\mathbf{y_1},\mathbf{y_2})  \eqdef \min\{\wt_H(\mathbf{z}) \mid \mathbf{z}\cdot G=\mathbf{y_1}-\mathbf{y_2}\},
\end{equation*} 
where $G$ is the $\abs{\cE}\times \abs{\In(\gamma)}$ submatrix of the extended global encoding matrix of $\cC$ which is defined in  Eq.~\eqref{eq:matG}. Noting that the rows of $G$ corresponding to the incoming links of $\gamma$  form an identity matrix, $d_{\cC}(\mathbf{y_1},\mathbf{y_2})$ is well-defined.   It is straightforward to verify that $d_{\cC}(\by_1,\by_2)$ is indeed a metric. 

Intuitively, $d_{\cC}(\by_1,\by_2)$ represents the minimum number of communication links in which an adversary must inject errors to transform the network output $\by_1$ into $\by_2$. In \cite{WeiXuGe23}, the distance  of a linear network computing code was defined  using $d_{\cC}(\cdot, \cdot)$, and it was shown in \cite[Lemma IV.1]{WeiXuGe23} 
that this definition is equivalent to the one provided in \eqref{eq:disdef}. Specifically, we have the following equality:   
\begin{align}\label{eq:altdist}
  d_{\min}(\cC,T,k) = \min \{ & d_{\cC}(\bx F,\bx'F) \mid \bx,\bx'\in \F_q^{sk} \textup{ and } \bx (T\otimes I_k)\neq\bx'(T\otimes I_k) \}.
\end{align}

Now, we can present the decoder. Let \[\cA\eqdef \set{\bx(T\otimes I_k); \bx \in \F_q^{sk}}\] 
be the set of all possible computing results.  For each $\ba \in \cA$, denote \[\Phi_{\ba}\eqdef \set{\bx F; \bx(T\otimes I_k)=\ba}.\] Given a received message $\tilde{\by}$, if there is a unique $\ba \in \cA$ such that $\Phi_{\ba}$ contains at least one vector $\by$ with $d_{\cC}(\by,\tilde{\by})\leq \tau$, the decoder then outputs  $\ba$ as the computing result; otherwise, it outputs an ``error". 

The following theorem justifies this decoding method.

\begin{thm} Let $\cC$ be a linear network code for a network $\cN$ with $d(\cC,T,k)\geq 2\tau+1$. If there are at most $\tau$ erroneous links in the network, then the decoder above outputs the correct computing result.
\end{thm}

\begin{IEEEproof} For a received message $\tilde{\by}$, since there are at most $\tau$ errors,  $\tilde{\by}$ can be written as $\tilde{\by}=\bx F +\bz G$ for some vectors $\bx \in \F_q^{sk}$ and $\bz \in \F_q^{\abs{\cE}}$ with $\wt_H(\bz)\leq \tau$. Let $\by=\bx F$, then 
\begin{equation}\label{eq:dec-1}
d_{\cC}(\by,\tilde{\by})\leq \tau.
\end{equation}
This shows that the correct  computing result $\ba = \bx(T\otimes I_k)$ satisfies the condition that $\Phi_{\ba}$ contains a vector $\by$ with $d_{\cC}(\by,\tilde{\by})\leq \tau$.

For another vector $\by'=\bx' F$ such that  $\bx' (T\otimes I_k)\neq\bx(T\otimes I_k)$, by the triangle inequality, we have that 
\[d_{\cC}(\by',\tilde{\by})\geq d_{\cC}(\by',\by)-d_{\cC}(\by,\tilde{\by}).\]
By \eqref{eq:altdist}, we have $d_{\cC}(\by,\by')\geq d_{\min}(\cC,T,k)$.
Hence, 
\begin{equation}\label{eq:dec-2}
d_{\cC}(\by',\tilde{\by})\geq d_{\cC}(\by,\by')-d_{\cC}(\by,\tilde{\by})\geq  d_{\min}(\cC,T,k) -d_{\cC}(\by,\tilde{\by}) \geq \tau+1.
\end{equation}
Eq.~\eqref{eq:dec-2} implies that for any $\ba'\neq \ba$,  $\Phi_{\ba'}$ does not contain any vector $\by'$ with $d_{\cC}(\by',\tilde{\by})\leq \tau$. Thus, the decoder can  output the correct computing result. 
\end{IEEEproof}

It is worth noting that the set $\Phi_{\ba}$ corresponds to a codeword in conventional coding theory, and the code $\cC$ encodes the computing result $\ba$ into $\Phi_\ba$.  The sink node receives an erroneous copy $\tby$ of a vector $\by$ of $\Phi_{\ba}$.  Since vectors from different $\Phi_\ba$'s have a large distance (with respect to $\cC$), the minimum distance decoder allows us to decode $\ba$, rather than $\by$, from the received $\tby$.

We now shift our focus to addressing link outages, i.e., links failing to transmit messages. Recall that if there is no error, the message transmitted by a link $e$ should be 
\begin{equation}\label{eq:encode}
u_e=\sum\limits_{d\in \In(\tail(e))} u_d k_{d,e}.
\end{equation}
Hence, if outages occur in a subset of links $\rho$,  in order to transmit the messages received by  $\tail(e)$,  the node can set $u_{d}=0$ for $d\in \In(\tail(e)) \cap \rho$ and then use \eqref{eq:encode} to encode the messages. In this way, the outages can be translated to the errors defined in Subsection~\ref{subsec:RNFCmodel}. Therefore, a code with $d_{\min}(\cC,T,k)\geq 2\tau+1$ is also robust to $\tau$ outages. 


In the following, we assume that the sink node is aware of the locations of outages. Under this assumption, similar to conventional codes with a minimum Hamming distance $d$ that can correct $d-1$ erasure errors, the network code $\cC$ can tolerate $d_{\min}(\cC,T,k) -1$ outages. 

\begin{thm} 
Let $\cC$ be a linear network code with  $d_{\min}(\cC,T,k)\geq \tau_o+1$. If there are at most $\tau_o$ outages in the network and their locations are known to the sink node, then $\cC$ is robust against these outages.
\end{thm}

\begin{IEEEproof}
Let $\rho_o$ denote the set of links where the outages occur, where  $\abs{\rho_o}  \leq \tau_o$. The received messages at the sink node can be expressed  as 
\[\tby = \bx F +\bz G + \bz_o,\] 
where $\bx \in \F_q^{sk}$, $\bz \in \F_q^{\abs{\cE}}$  is an imaginary error vector matching $\rho_o$, and $\bz_o \in \set{0, \star}^{\abs{\In(\gamma)}}$ is an indicator vector with  the symbol $\star$ representing an outage in the incoming links of $\gamma$. We define $\star + x =\star$ for all $x\in \F_q$. The nonzero entries of $\bz$ are chosen  such that  for every $d\in \rho_o\backslash \In(\gamma)$ the message received by $\head(d)$ is $\tilde{u}_d=0$.

Now, suppose to the contrary that there is another $\bx'\in \F_q^{sk}$ and $\bz'\in \F_q^{\abs{\cE}}$ matching $\rho_o$ such that 
\[\bx \cdot(T\otimes I_k) \neq \bx' \cdot(T\otimes I_k)\]
and
\[\bx F+\bz G+\bz_o=\bx' F +\bz' G+\bz_o.\]
Noting that $G$ contains an $\abs{\In(\gamma)}\times \abs{\In(\gamma)}$ identity matrix, we have 
\[(\bx-\bx')F = (\bu'-\bu )G\]
for some vectors $\bu,\bu'\in \F_q^{\abs{\cE}}$, both of which match $\rho_o$.
By \eqref{eq:disdef}, the size of the support of $\bu'-\bu$ is at least $d_{\min}(\cC,T,k)$. However, since both $\bu$ and $\bu'$ match $\rho_o$, then the support of $\bu'-\bu$ is contained in $\rho_o$. It follows that $\abs{\rho_o}\geq d_{\min}(\cC,T,k)\geq \tau_o+1$, which contradicts  the assumption that $\abs{\rho_o}\leq \tau_o$.
\end{IEEEproof}

The proof of the lemma above leads to the following decoder for the outages.  Let $\cA$
be the set of all possible computing results.  For each $\ba \in \cA$ and a subset $\rho \subseteq \cE$, denote \[\Phi_{\ba,\rho}\eqdef \set{\bx F+ \bz G +\bz_\rho; \bx(T\otimes I_k)=\ba, \ \bz \textup{ matching  }\rho},\]
where $\bz_\rho \in \set{0,\star}^{\In(\gamma)}$  indicates the links of $\In(\gamma)\cap \rho$.   
Then for a received message $\tby \in (\F_q \cup \set{\star})^{\In(\gamma)}$ and the set of outage locations $\rho_o$, the proof above shows there is a unique $\ba \in \cA$ such that $\tby \in \Phi_{\ba,\rho_o}$.

\begin{rmk}\label{rmk.erasure}
If the outages occur only in the incoming links of $\gamma$, then the message received at the sink node, $\tby=\bx \cdot F+\bz_o$, is  an erroneous copy of the vector $\by = \bx\cdot F $ with  $\abs{\rho_o}$ erasures. In this case, we can use the Hamming metric to design a simpler decoder. Let $\ba\in \cA$ such that $\by \in \Phi_\ba$. For any $\by' \in \Phi_{\ba'}$ with $\ba'\neq \ba$, the Hamming distance between $\by$ and $\by'$ satisfies 
\[d_H(\by, \by') \overset{(*)} {\geq} d_{\cC}(\by,\by')\overset{(**)}{\geq} d_{\min}(\cC,T,k) >\abs{\rho_o}, \]
where the inequality $(*)$ holds because   $d_{\cC}(\by,\by')  = \min\{\wt_H(\mathbf{z})\mid \mathbf{z}\cdot G=\mathbf{y}-\mathbf{y}'\}$ and the rows of $G$ corresponding to the incoming links of $\gamma$  form an identity matrix, and the inequality $(**)$ follows from Eq.~\eqref{eq:altdist}. Hence, for the received message $\tby$, there is a unique $\ba\in \cA$ such that $\Phi_\ba$ contains a vector $\by$ that matches $\tby$ on all the non-star components. The decoder  then outputs this $\ba$ as the computing result.  
\end{rmk}

\section{Computing the Sum Function Against Errors}\label{Sec:sum_network}

The problem of network function computation, particularly the sum function, has been extensively studied in the literature \cite{2004networkcodes,2008Ramamoorthy,2013RaiDas,2013RamamoorthyLangberg,2018Tripathy,2012RD,2010ShenviDey,RaiDeyShe2010} under the assumption that there are no errors in the network. When there is only one sink node, linear network coding can achieve the computing capacity for an arbitrary directed acyclic network and the sum function. This coding scheme is derived from the equivalence between  computing the sum and multicasting source messages in the reverse network.  For a network $\cN$, its \emph{reverse network} $\cN^r$ is obtained by reversing the direction of the links and interchanging the roles of source nodes and sink nodes. In the scenario where no error links occur, computing the sum in $\cN$ is equivalent to multicasting source messages to all the sink nodes in the reverse network $\cN^r$. Specifically, if there is a 
linear network code for $\cN^r$ that can multicast the  messages generated at the source sink to all the sink nodes with an information rate $h$, then by reversing the network and dualizing all local behaviors, one can obtain a linear network code for $\cN$ that computes the sum of sources with a computing rate $h$, as demonstrated in  \cite[Theorem~5]{2004networkcodes}. Conversely, the same principle applies in the reverse direction. By leveraging this connection, it can be
shown that the cut-set bound on the computing capacity can
be attained for a generic network by linear network coding. 

However, if there are errors in the links, this equivalence no longer holds. As shown in \cite[Example IV.1]{WeiXuGe23}, the dual of a linear network code that is resilient to a single error in the multicast problem cannot correctly compute the sum in the reverse network when an error occurs. In this section, we will show that although this equivalence does not hold, the cut-set bound on the robust computing capacity  still can be achieved by linear network coding. Our approach is to modify the dual of a linear network code for the multicast problem without error tolerance to obtain a linear network code, denoted by $\cC$, capable of computing the sum with a certain level of error tolerance, albeit at a lower computing rate of $k$. Surprisingly,  the distance of this code attains the Singleton bound, that is, \[d_{\min}(\cC,\bOne,k)=\min_{C\in \Lambda(\cN)} \set*{\abs{C} -k+1}.\]

The proposed coding scheme consists of the following two parts:

\begin{enumerate}
  \item {\bf Internal coding}. In this part, we design the transfer matrix of $\cC$, which describes the encoding functions at all internal nodes, i.e., the nodes in $\cV \backslash (S\cup\set{\gamma})$. Consider a multicast problem in the reverse network without link errors. It is well known  that if the field size is larger than the number of sink nodes, then  a linear network code exists, which can multicast messages at a rate of $h\eqdef \min_{C\in \Lambda(\cN)} \set{\abs{C}}$,  e.g.,  see \cite{JagSanChoEffEgnJaiTol05}. Let $K$ denote the transfer matrix of this code. We use the transpose $K^\top$ as the transfer matrix of the computing code $\cC$. Note that the matrix $G$ of $\cC$ can then be determined  via \eqref{eq:matG}.
  \item {\bf External/source coding}. In this part, we  carefully design the source encoding matrices $B_i$ such that the  distance of the proposed  network coding scheme $\cC$ achieves the Singleton-like bound. Recall that 
  \[d_{\min}(\cC,T,k)= \min\set{\abs{\rho};\Phi\cap\Delta(\rho)\neq\varnothing},\]
  where
  \[
  \Phi= \set{\bx\cdot F  ; \bx(T\otimes I_k)\neq\mathbf{0},\  \bx\in\F_q^{sk}}
\]
and
\[\Delta(\rho) = \set{\mathbf{z}\cdot G|\  \mathbf{z}\in\F_q^{|\mathcal{E}|} \textup{ matching the error pattern }\rho}.\]
  Given a computing rate $k$ which is smaller than $h$, we first choose a subspace $W$ of $\F_q^{\abs{\In(\gamma)}}$ which intersects each $\Delta(\rho)$ trivially, where $\abs{\rho}\leq h -k$. Noting that  $F$ is determined by $K$ and $B_i$'s via \eqref{eq:matF}, which in turn determines $\Phi$, 
    we then design the source encoding $B_i$ to ensure that  $\Phi$ is contained in $W$. In this manner, the condition in \eqref{eq:disdef} is fulfilled, and so, the distance of code achieves the upper bound.  
\end{enumerate}

The internal coding part is straightforward, while the source coding part is more intricate. We first use an example to illustrate our approach.

\begin{figure}
  \centering
  \includegraphics[width=12cm]{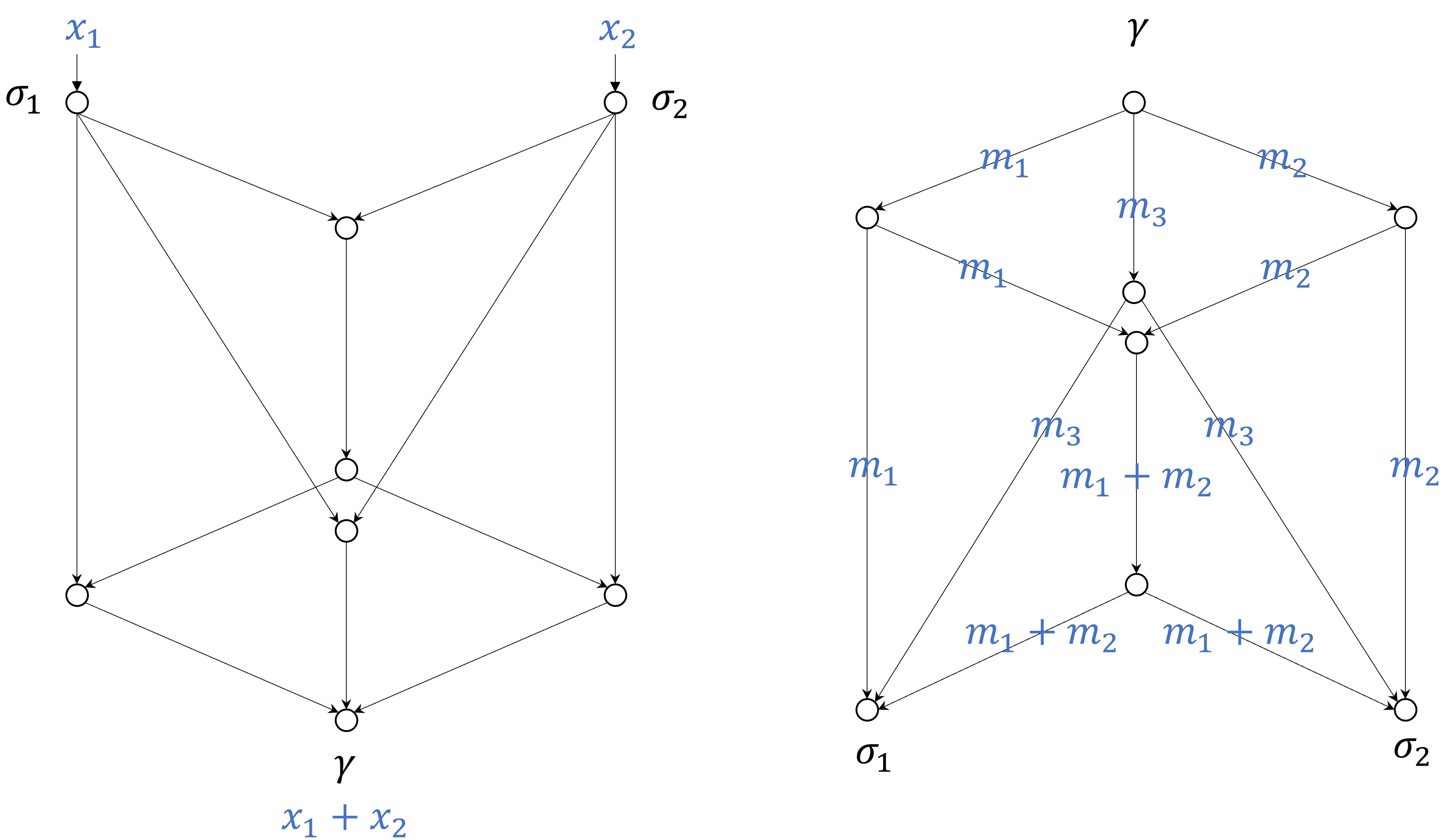}
  \caption{The network on the left is a sum-network, where each source $\sigma_i$ generates a message $x_i$ and the sink node wants to compute the sum $x_1+x_2$. The network on the right is a multicast network, along with a coding scheme which achieves the maximum communication rate $3$. }\label{fig:example-1}
\end{figure}

\begin{figure}
  \centering
  \includegraphics[width=7cm]{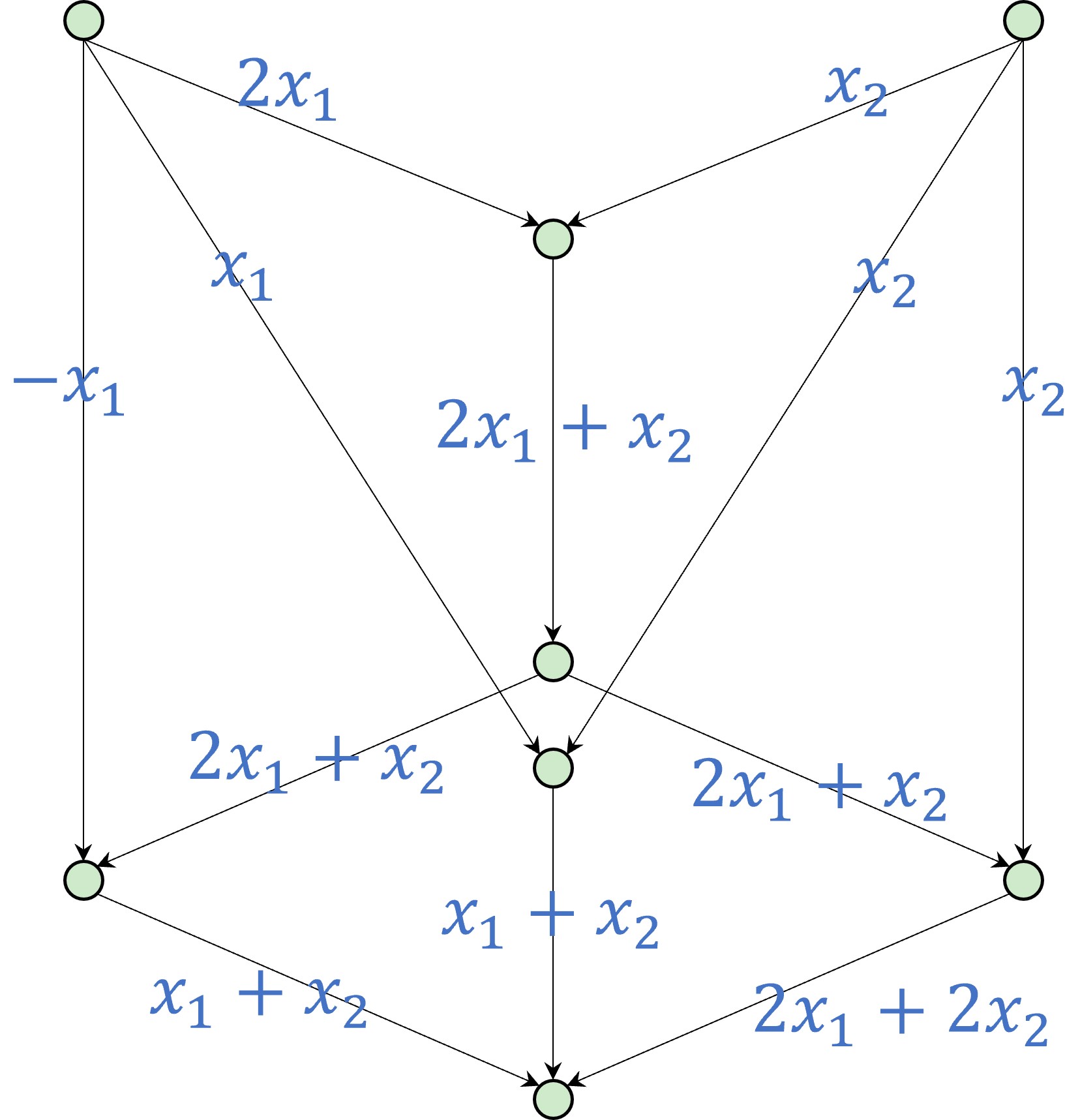}
  \caption{A coding scheme designed to compute  $x_1+x_2$ over $\F_q$, where $q$ is odd.}\label{fig:srcenc}
\end{figure}

\begin{figure}
  \centering
  \includegraphics[width=7.5cm]{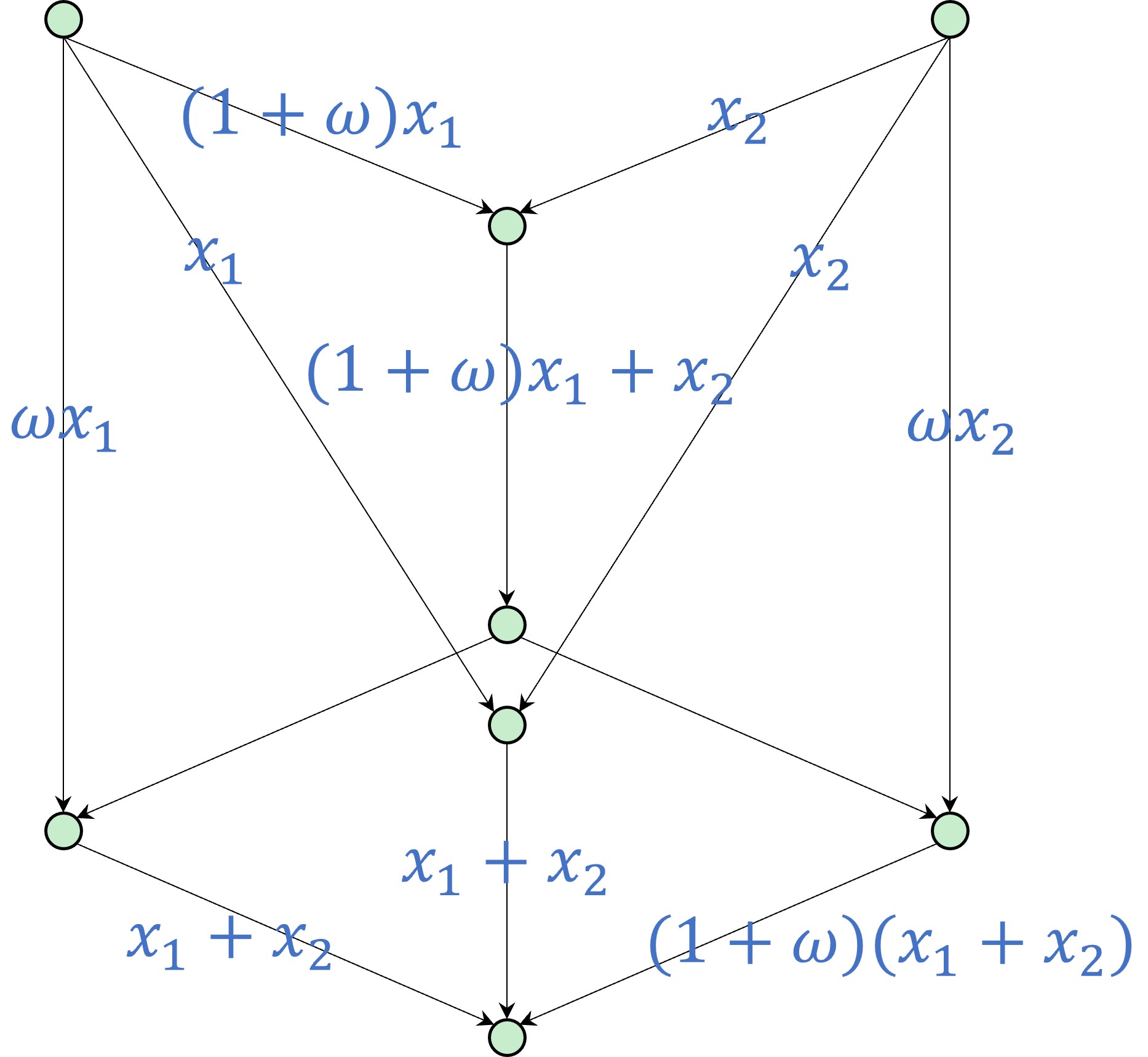}
  \caption{A coding scheme designed to compute  $x_1+x_2$ over $\F_q$, where $q$ is even.  }\label{fig:srcenc-2}
\end{figure}

\begin{example}\label{ex:toy}
Consider the network shown on the left of Fig.~\ref{fig:example-1}. It has two source nodes $\sigma_1,\sigma_2$ and a sink node $\gamma$. The min-cut capacity between each source $\sigma_i$ and $\gamma$ is $3$. Let $q\geq 3$ be a prime power  and $\F_q$ be a finite field of size  $q$. We are going to design a linear network coding scheme over $\F_q$ which can compute the sum $x_1+x_2$ at  a rate of $1$ even when one link  is corrupted.

Consider the reverse network shown on the right of Fig.~\ref{fig:example-1}, together with a mutlcast coding scheme with an information rate of $3$. Note that in this scheme every internal node simply adds up the received messages and transmits the result to the downstream nodes. Thus, its transfer matrix $K$ is a binary matrix and $K_{d,e}=1$ if and only if $\head(d)=\tail(e)$. In the sum network, we also require every internal node to add up the received messages and transmit the result to the downstream nodes, so that the transfer matrix is $K^\top$. Once the transfer matrix is fixed, the submatrix $G$ in the extended global encoding matrix $\tF$ is determined. In this instance, $G$ consists of $12$ rows, where the vectors ${(1,0,0),(0,1,0),(0,0,1),(1,0,1)}$  appear three times each.

For the source encoding, we proceed with two cases. First assume that $q$ is odd. In this case, we encode $x_1$ into $-x_1$, $2x_1$ and $x_1$ at $\sigma_1$, and  encode $x_2$ into $x_2$, $x_2$ and $x_2$ at  $\sigma_2$. The entire coding scheme $\cC$ is illustrated  in Fig.~\ref{fig:srcenc}. Then the global encoding matrix 
\[F =\begin{pmatrix}  1 & 1 & 2 \\ 1 & 1 & 2\end{pmatrix}.\]
If there are no link errors, the sink node $\gamma$ receives $\by= (x_1,x_2)\cdot F = (x_1+x_2,x_1+x_2,2(x_1+x_2))$, and so, it can compute the sum $x_1+x_2$. Furthermore, note that 
\[
  \Phi= \set{(x_1,x_2)\cdot F  ; x_1+x_2\neq 0,\  \bx = (x_1,x_2)\in\F_q^{2}} = \set{x\cdot (1,1,2); x\in \F_q\backslash \set{0}}
\]
and that the vector $(1,1,2)$ is not contained in any subspace spanned by at most two vectors of \[\set{(1,0,0),(0,1,0),(0,0,1),(1,0,1)}.\] It follows that for any $\rho \subset \cE$ with $\abs{\rho}=2$, we have that
\[\Phi \cap \Delta(\rho)= \varnothing. \] 
Hence, the proposed coding scheme has distance at least $3$, and so, can tolerate one link error.

For the case where $q$ is an even prime power, let $\omega$ be a primitive element of $\F_q^*$. We encode $x_1$ into $\omega x_1$, $x_1$ and $(1+\omega)x_1$ at $\sigma_1$, and  encode $x_2$ into $x_2$, $x_2$ and $\omega x_2$ at  $\sigma_2$. The entire coding scheme $\cC$ is illustrated  in Fig.~\ref{fig:srcenc-2}. Then the global encoding matrix 
\[F =\begin{pmatrix}  1 & 1 & 1+\omega \\ 1 & 1 & 1+\omega\end{pmatrix}.\]
By the same argument as above, one can show that the distance of the code is $3$.

In the following, we give a decoding algorithm, which is simpler than the minimum distance decoder presented in Section~\ref{Sec:mindisdec}. Suppose that there is a link error and $q$ is odd. The sink node $\gamma$  gets
$\tby= \by+\bz$ where $\by=(x_1,x_2)\cdot F = (x_1+x_2,x_1+x_2,2(x_1+x_2))$ and $\bz$ is contained in one of the following spaces:
\[ \angenv{(1,0,0)}, \angenv{(0,1,0)}, \angenv{(0,0,1)}, \angenv{(1,0,1)}. \]
Assume that $\tby=(\ty_1,\ty_2,\ty_3)$ and $\by=(y_1,y_2,y_3)$. 
Consider the following decoder:
\[
\cD(\tby)=\begin{cases}
\ty_1 & \text{if  $2\ty_1=\ty_3$}, \\
 \ty_2 & \text{otherwise.} \\
\end{cases}
\]
If $2\ty_1=\ty_3$, then $\bz$ must be contained in $\angenv{(0,1,0)}$. Thus, $\ty_1=y_1=x_1+x_2$. If $2\ty_1\neq \ty_3$, necessarily  $\angenv{(0,1,0)}$ does not contain $\bz$, and so, $\ty_2=y_2=x_1+x_2$.  Hence, $\cD(\tby)$ can decode the sum $x_1+x_2$. 
\end{example}

In general, the source encoding process can be outlined as follows. Once the internal encoding is fixed, the submatrix $G$ in the extended global encoding matrix $\tF$ is determined. To achieve the Singleton bound, we aim to find a $k\times \abs{\In(\gamma)}$ matrix\footnote{In Example~\ref{ex:toy}, we have $D=(1,1,2)$.} $D$ such that
 \begin{enumerate}
   \item[(P1)] $D$ contains a $k\times k$ identity matrix $I$ as submatrix;
   \item[(P2)] the row space of $D$ intersects trivially with any space spanned by at most $h-k$ rows of $G$, where $h=\min_{C\in \Lambda(\cN)} \set{\abs{C}}$.
 \end{enumerate}
 Our internal encoding allows us to design source encoding such that
\begin{equation}\label{eq:receive}
\bx_S \cdot F = \parenv*{ \sum_{i=1}^s{\bx_i}} \cdot D,
\end{equation}
This equality, combined with properties (P1) and (P2), ensures that the proposed code can achieve the Singleton-like bound.

To guarantee  the existence of the matrix $D$, we introduce  the following lemma.

\begin{lem}\label{lm:ineqqbinom}
Let $h,k$ and $E$ be fixed positive integers such that $0< h-k \leq E$. Let $q$ be a  sufficiently large prime power. Then we have that
  \begin{equation*}
   \parenv*{\sbinom{h}{k}_q-q^{k(h-k)}} \binom{E}{h-k}  < \sbinom{h}{k}_q,
  \end{equation*}
  where $\sbinom{h}{k}_q \eqdef \prod\limits_{i=0}^{k-1} \frac{q^h-q^i}{q^k-q^i}$ is the Gaussian coefficient.
\end{lem}

\begin{IEEEproof}
Let 
\begin{align*}
A & \eqdef (q^h-1)(q^h-q)\cdots (q^h-q^{k-1}), \\
B & \eqdef (q^h-q^{h-k}) (q^h-q^{h-k+1})\cdots (q^h-q^{h-1}),\\
C & \eqdef (q^k-1)(q^k-q)\cdots (q^k-q^{k-1}).
\end{align*}
Then 
\[ A = q^{kh}- \parenv*{\sum_{i=0}^{k-1} q^i} q^{h(k-1)} + O(q^{k-1+k-2}q^{h(k-2)} ) = q^{kh} - O(q^{kh-(h-k)-1}) \]
and
\[ B =  q^{kh}-q^{kh-1}+O(q^{kh-2}). \]
It follows that  
\[A -B =q^{kh-1}+O(q^{kh-2}).\]
Hence, for fixed $k,h,E$ and sufficiently large $q$, we have that 
\[(A-B)\binom{E}{h-k} < A .  \]
Dividing both sides by $C$ and noting that $\sbinom{h}{k}_q =A/C$ and $q^{k(h-k)}=B/C$, we then get
\[\parenv*{\sbinom{h}{k}_q-q^{k(h-k)}} \binom{E}{h-k}  < \sbinom{h}{k}_q.\]
\end{IEEEproof}

\begin{thm}\label{thm:Sbndtight}
Let $\cN$ be an arbitrary directed acyclic network and $k$ be a positive integer such that $k \leq \min_{C \in \Lambda(\cN)} \set{\abs{C}}$. If $q$ is sufficiently large,
then there is a linear network code $\cC$ over $\F_q$ which can compute the sum of the source messages with
\[d_{\min}(\cC,\bOne, k) = \min_{C \in \Lambda(\cN)} \set{\abs{C}-k+1}.\]
\end{thm}

\begin{IEEEproof} 
Denote  $h\eqdef \min_{C\in \Lambda(\cN)} \set{\abs{C}}$.  We assume that $\abs{\Out(\sigma_i)}=h$ for each source node $\sigma_i \in S$ and $\abs{\In(\gamma)}=h$ for the sink node $\gamma$, by adding auxiliary source and sink nodes and connecting each of them to the original one by $h$ links. Let $\cN^r$ denote the reverse network of $\cN$. Consider a multicast problem over $\cN^r$, where each node $\sigma_i$ demands $h$ messages generated at the node $\gamma$. We use a superscript $r$ to distinguish the notations for $\cN^r$ from the ones for $\cN$. For example, $\cE^r$ denotes the set of links of $\cN^r$; $\Out^r(\gamma)$ denotes the set of outgoing links of $\gamma$ in $\cN^r$, which can be obtained by reversing each link of $\In(\gamma)$.

 Since
\[\min_{C\in \Lambda(\cN^r)} \set{\abs{C}} =  \min_{C\in \Lambda(\cN)} \set{\abs{C}}=h,\]
when $q>s$, there is a linear solution to this multicast problem. In other words, there are matrices    $B = ( k_{i,e} )_{ i \in [h], e \in \cE^r}$ and $K= (k_{d,e})_{d \in \cE^r, e\in \cE^r}$ over $\F_q$, where $k_{i,e}=0$ if $e$ is not an outgoing edge of  $\gamma$ in $\cN^r$ and $k_{d,e}=0$ if $e$ is not an outgoing edge of $\head(d)$ in $\cN^r$, such that
\begin{equation}\label{eq:mulcastsol}
B \cdot (I-K)^{-1} \cdot \begin{pmatrix} A_{\In^r(\sigma_1)}^\top & A_{\In^r(\sigma_2)}^\top & \cdots & A_{\In^r(\sigma_s)}^\top \end{pmatrix}  = (F_1,F_2,\ldots, F_s)
\end{equation}
for some full rank matrices $F_i \in \F_q^{h\times h}$, where $I$ is the $\abs{\cE^r}\times \abs{\cE^r}$ identity matrix.

Now, we direct our attention to the network $\cN$. We use the transpose of $K$ to give the local encoding coefficients for each node $u \in \cV \backslash (S \cup \set{\gamma})$, namely, for each link $e \notin  \cup_{\sigma_i \in S} \Out(\sigma_i)$, let
\[ \bu_e = \sum_{d\in \In(\tail(e))}k_{e',d'}\bu_d,\]
where $e'$ and $d'$ are the corresponding links of $e$ and $d$ in $\cE^r$, respectively. 

We now present the local encoding coefficients for each source node $\sigma_i$ such that the distance of the proposed code attains the Singleton-like bound, i.e., $d_{\min}(\cC,\bOne, k) = \min_{C \in \Lambda(\cN)} \set{\abs{C}-k+1}=h-k+1$.

Recall that
\[\Delta(\rho) = \set*{\mathbf{z}\cdot G|\  \mathbf{z}\in\F_q^{|\mathcal{E}|} \textup{ matching the error pattern }\rho}\]
can be treated as a subspace of $\F_q^h$ that is generated by $\abs{\rho}$ rows of $G$, where \[G = (I-K^\top)^{-1}A_{\In(\gamma)}^\top. \]
Given an $(h-k)$-dimensional subspace $U$ of $\F_q^{h}$, the number of $k$-dimensional subspaces which intersect $U$ trivially is
\[ \frac{(q^h-q^{h-k})(q^h-q^{h-k+1})\cdots (q^h-q^{h-1})}{(q^k-1)(q^k-q)\cdots (q^k-q^{k-1})  } =q^{k(h-k)}.\]
Since $q$ is sufficiently large, according to Lemma~\ref{lm:ineqqbinom}, we have that 
\begin{equation*} 
\parenv*{\sbinom{h}{k}_q-q^{k(h-k)}} \binom{\abs{\cE}}{h-k}  < \sbinom{h}{k}_q.
\end{equation*}
It follows that there is a $k$-dimensional subspace $W$ of $\F_q^h$ such that
\begin{equation}\label{eq:Wdisjt}
W \cap  \Delta(\rho) =\set{\Zero},
\end{equation}
for any $\rho \subseteq \cE$ with $\abs{\rho}\leq h-k$.
Let $D\in \F^{k\times h}_q$  be a matrix whose rows form a basis of $W$ such that it contains the $k \times k$ identity matrix $I_k$ as a submatrix.

Noting that the support set of $B$ is contained in the support set of  $A_{\Out^r(\gamma)}$, let $\tilde{B}$ be an $h\times h$ matrix over $\F_q$ such that $B=\tilde{B} \cdot A_{\Out^r(\gamma)}$. Since $F_i$'s are invertible, by \eqref{eq:mulcastsol}, $\tilde{B}$ is also invertible. For each $i=1,2,\ldots,s$,  let 
\[F_i'\eqdef F_i^\top \cdot (\tilde{B}^\top)^{-1}\in \F_q^{h\times h},\]
and
\begin{equation}\label{eq:souenc}
E_i\eqdef D \cdot (F_i')^{-1} \cdot A_{\Out(\sigma_i)}.
\end{equation}
Then $E_i \in \F_q^{k\times \abs{\cE}}$. Moreover, its columns which are indexed by the edges not in $\Out(\sigma_i)$ are all-zero vectors. Hence, each $E_i$ can be used to describe  local encoding coefficients at  $\sigma_i$.

Let $\cC$ be the coding scheme described by $E_i$'s and $K^\top$. Now, we show that $\cC$ can compute the sum function and $d_{\min}(\cC,\bOne, k) =h-k+1$. 
We transpose both sides of \eqref{eq:mulcastsol}. Noting that $A_{\In^r(\sigma_i)}=A_{\Out(\sigma_i)}$, $A_{\Out^r(\gamma)}=A_{\In(\gamma)}$ and $B^{\top}=  A_{\Out^r(\gamma)}^\top \tilde{B}^\top$, we have that
\begin{equation}\label{eq:transpose}
 \begin{pmatrix} A_{\Out(\sigma_1)} \\ A_{\Out(\sigma_2)} \\ \vdots \\ A_{\Out(\sigma_s)}\end{pmatrix} \cdot (I-K^\top)^{-1} \cdot A_{\In(\gamma)}^\top  \cdot \tilde{B}^\top = \begin{pmatrix} F_1^\top \\ F_2^\top \\ \vdots \\ F_s^\top \end{pmatrix}.
\end{equation}
Multiplying both sides of \eqref{eq:transpose} with $ (\tilde{B}^\top)^{-1}$, we have that
\begin{equation}\label{eq:transpose-1}
 \begin{pmatrix} A_{\Out(\sigma_1)} \\ A_{\Out(\sigma_2)} \\ \vdots \\ A_{\Out(\sigma_s)}\end{pmatrix} \cdot (I-K^\top)^{-1} \cdot  A_{\In(\gamma)}^\top = \begin{pmatrix} F_1' \\ F_2' \\ \vdots \\ F_s' \end{pmatrix}.
\end{equation}

Then for the network code $\cC$, its global encoding matrix
\begin{align*}
F & = \begin{pmatrix} E_1 \\ E_2\\ \vdots \\ E_s \end{pmatrix} \cdot (I-K^\top)^{-1}\cdot A_{\In(\gamma)}^{\top} \\
  & \overset{\eqref{eq:souenc}}{=} \begin{pmatrix} D \cdot (F_1')^{-1} & & & \\ &  D \cdot (F_2')^{-1} & & \\ & & \ddots & \\ & & & D \cdot (F_s')^{-1} \end{pmatrix} \cdot  \begin{pmatrix} A_{\Out(\sigma_1)} \\ A_{\Out(\sigma_2)} \\ \vdots \\ A_{\Out(\sigma_s)}\end{pmatrix} \cdot (I-K^\top)^{-1} \cdot  A_{\In(\gamma)}^\top \\
   & \overset{\eqref{eq:transpose-1}}{=}     \begin{pmatrix} D \cdot (F_1')^{-1} & & & \\ &  D \cdot (F_2')^{-1} & & \\ & & \ddots & \\ & & & D \cdot (F_s')^{-1} \end{pmatrix} \cdot \begin{pmatrix}  F_1' \\  F_2' \\ \vdots \\  F_s' \end{pmatrix} = \begin{pmatrix}  D \\ D \\ \vdots \\  D \end{pmatrix}.
\end{align*}
Therefore, for a vector $\bx_s =(\bx_1,\bx_2,\ldots,\bx_s) \in \F_q^{sk}$, we have that
\begin{equation}\label{eq:cmptF}
\bx_S \cdot F =  \sum_{i=1}^s (\bx_i \cdot D)=\parenv*{\sum_{i=1}^s \bx_i} \cdot D.
\end{equation}
Since $D$ contains the identity matrix $I_k$, the sink node can receive $\sum_{i=1}^s \bx_i$, i.e., the coding scheme can compute the sum $k$ times. Furthermore,
\[
  \Phi= \set*{\bx_S \cdot F  ;  \bx_S=(\bx_1,\bx_2,\ldots,\bx_s)  \in\F_q^{sk}, \ \sum_{i=1}^s\bx_i \neq\mathbf{0}}\overset{\eqref{eq:cmptF}}{=}\set{\bx \cdot D ; \bx\in \F_q^k\backslash\set{\Zero}},
\]
which consists of  the nonzero vectors of the subspace $W$. Hence, for any $\rho \subseteq \cE$ with $\abs{\rho}\leq h-k$, by \eqref{eq:Wdisjt}, we have that
\[\Phi \cap \Delta(\rho) = \varnothing,\]
namely,
\[d_{\min}(\cC,\bOne, k) \geq h-k+1.\]
\end{IEEEproof}

\section{Computing the Identity Function Against Errors}\label{Sec:genfun}

In this section, we examine  the case of $l=s$ and show that the Singleton-like bound can still be achieved. As discussed in Section~\ref{sec:prelim}, it suffices to focus on the identity function. In this scenario, the computing problem involves transmitting multiple source messages to a single sink node.  In the error-free model, the achievability of the cut-set bound on communication capacity is established by considering an augmented network, where an auxiliary source node is connected to each original source node $\sigma_j$ by $R_i$ links. Here, the $R_i$'s represent information rates that satisfy the condition imposed by the cut-set bound.  A solution that only uses message routing for the unicast problem in this augmented network yields a solution for the multi-message transmitting problem in the original network, as shown in \cite[Theorem 4.2]{lehman2004}.   However, for the error correction  problem, this approach may fail: in an error-correcting scheme for unicasting in the augmented network, the outgoing messages from different source nodes may  be dependent as they originate  from  the auxiliary node, whereas the topology of the original network requires   messages from different source nodes to be independent.

Our proof utilizes the ideas and concepts  presented in \cite{2008Zhang,guang2014linear}, where  linear network error-correction codes  for the multicast problem were studied. We begin by introducing some notation.  Let $\rho,\rho'\subseteq\mathcal{E}$ be two error patterns. We say $\rho'$ \emph{dominates} $\rho$ if $\Delta(\rho)\subseteq\Delta(\rho')$ for any linear network code. This relation is denoted by $\rho\prec\rho'$. Let $\Rank(\rho)$ be the \emph{rank} of an error pattern $\rho$, which is defined as
\[\Rank(\rho)\eqdef \min\set{\abs{\rho'} ;\rho\prec\rho'}.\]
For a positive integer $\delta$, let $R(\delta)$ be a collection of error patterns which is defined as
\[R(\delta)\eqdef \set{\rho ;
\abs{\rho}=\Rank(\rho)=\delta}.\]

When $T$ is the identity matrix, the Singleton-like bound in Theorem~\ref{thm:Sinbnd} reads: 
\[d_{\min}(\cC,I,k)\leq\min\limits_{C\in\Lambda(\cN)}\set{|C|-k|I_C|+1}.\]
Denote $\delta\eqdef \min_{C\in\Lambda(\cN)}\set{|C|-k|I_C|}$. In order to prove that this bound is achievable, we need to show that there exists a linear network code with $d_{\min}(\cC,I,k)>\delta$. 
Due to the definition of $d_{\min}(\cC,I,k)$, it suffices to prove that for every $\rho\in R(\delta)$, $\Phi\cap\Delta(\rho)=\varnothing.$ The proof can be divided into two steps. First, we will show that for every $\rho\in R(\delta)$, there are $sk+\delta$ edge-disjoint  paths, where $k$ paths are from $\sigma_i$ (for each $i \in [s]$) to $\gamma$, and $\delta$ paths are from $\rho$ to $\gamma$.  Second, for every $\rho\in R(\delta)$, we define a dynamic set $CUT_{\rho}$ and update the global encoding vectors of the edges in this set until all global encoding vectors have been updated.

Let $\cN$ be a directed acyclic network.  For an error pattern $\rho\subseteq \cE$, we construct a new network, denoted by $\cN_{\rho}$, which is obtained by adding a new  node $\sigma_\rho$ and creating a new link $e'=(\sigma_\rho,\head(e))$ for each $e \in \rho$.

\begin{lem}\label{lem:Rank(rho)=mincut} The rank of  an error pattern $\rho\subseteq \cE$ is equal to the size of the minimum cut between $\sigma_\rho$ and $\gamma$ in the network $\cN_\rho$.
\end{lem}
\begin{IEEEproof}
 For an arbitrary linear network code of $\cN$, we can define a linear network code of $\cN_\rho$ by letting $k_{e',d}=k_{e,d}$ for all $e\in \rho$ and $d\in \Out(\head(e))$ and keeping all the other local encoding coefficients. Let $G$ and $G'$ be the submatrices of the extended global encoding matrices for $\cN$ and $\cN_\rho$, respectively. Then the row labeled by $e'$ in $G'$  is equal to the row labeled by $e$ in $G$. It follows that $\Delta(
 \rho)=\Delta(\rho')$, where $\rho'\eqdef \set{e' ; e \in \rho}$. Let $C_{\sigma_\rho,\gamma}$ be a minimum cut between $\sigma_\rho$ and $\gamma$. Since every path from $\sigma_\rho$ to $\gamma$ must pass through $C_{\sigma_\rho,\gamma}$, the row labeled by $e'$ must be a linear combination of the rows of $G'$ that are labeled by the links in $C_{\sigma_\rho,\gamma}$. Hence,
    \[\Delta(\rho)=\Delta(\rho')\subseteq\Delta(C_{\sigma_\rho,\gamma}),\]
    which implies that $\Rank(\rho)\leq \abs{C_{\sigma_\rho,\gamma}}$.

    On the other hand, for any linear network code, we have 
    \begin{align*}
    \Rank(\rho) &=\min\set{|\rho'| ; \rho\prec\rho'}\\
    &\geq\min\set{\dim(\Delta(\rho')) ; \rho\prec\rho'}\\
    &\geq \dim(\Delta(\rho)).
    \end{align*}
    Take  an arbitrary set of $|C_{\sigma_\rho,\gamma}|$ edge-disjoint paths from $\sigma_\rho$ to $\gamma$. Construct a linear network code by setting the local encoding coefficient $k_{d,e}=1$ if $d,e$ belongs to the same path, and $k_{d,e}=0$ otherwise. For this particular linear code, it is obvious that $\dim(\Delta(\rho))=|C_{\sigma_\rho,\gamma}|$. Therefore, we have $\Rank(\rho)\geq|C_{\sigma_\rho,\gamma}|$.
\end{IEEEproof}

\begin{lem}\label{lem:cutrho}
Let $\rho \in R(\delta)$ be an error pattern. In the network $\cN_{\rho}$, we add a new source node $\sigma'$, together with $\delta$ links from $\sigma'$ to $\sigma_\rho$, and $k$ links from $\sigma'$ to $\sigma_i$ for each $1\leq i\leq s$. Then the size of the minimum cut between $\sigma'$ and $\gamma$ in this new network is equal to $sk+\delta$. 
\end{lem}

\begin{IEEEproof}
Since $\abs{\Out(\sigma')}=sk+\delta$, the size of the minimum cut between $\sigma'$ and $\gamma$ is at most $sk+\delta$. To show it is at least this number, we consider an arbitrary cut $C$  separating $\sigma'$ and $\gamma$. Let $C_1\eqdef C \cap \Out(\sigma')$ and $C_2\eqdef  C\backslash C_1$. We proceed with the following cases. 
\begin{enumerate}
\item If $I_{C_2}=\varnothing$ and $C_2$ is not a cut between  $\gamma$ and $\sigma_\rho$, then it must be the case that $C_1=\Out(\sigma')$. Hence, $\abs{C} \geq \abs{C_1}\geq \abs{\Out(\sigma')}=sk+\delta$. 
\item If $I_{C_2}=\varnothing$ and $C_2$ is a cut between  $\gamma$ and  $\sigma_\rho$, then $\cup_{i=1}^s {\In(\sigma_i)}\subseteq C_1$, and $\abs{C_2}\geq \Rank(\rho)=\delta$ (by Lemma~\ref{lem:cutrho}).  It follows that $\abs{C} =\abs{C_1}+\abs{C_2}\geq sk+\delta$. 
\item If $I_{C_2}\neq \varnothing$, then $C_2$ is a cut of the original network $\cN$. It follows that $\abs{C_2}\geq k\abs{I_{C_2}}+\delta$, as  $\delta=\min_{C\in\Lambda(\cN)}\set{|C|-k|I_C|}$. Note that $C_2$ only separates $\gamma$ from the source nodes in $I_{C_2}$.  Then $\cup_{\sigma_i \in S \backslash {I_{C_2}}} \In(\sigma_i) \subseteq C_1$, and so, $\abs{C_1}\geq sk - \abs{I_{C_2}}k$. Hence, $\abs{C}=\abs{C_1}+\abs{C_2}\geq sk+\delta$.  
\end{enumerate}
\end{IEEEproof}

Using this lemma, we can prove the following result.

\begin{cor}\label{cor:edge-disjoint_path}
    For every error pattern $\rho\in R(\delta)$, there are $(sk+\delta)$  edge-disjoint paths, where  $\delta$ paths are from $\rho$ to $\gamma$, each starting from a link in  $\rho$, and $k$ paths are from $\sigma_i$ to $\gamma$ for each $1\leq i\leq s$.
\end{cor}
\begin{IEEEproof} 
    For every $\rho\in R(\delta)$, consider the network in Lemma~\ref{lem:cutrho}. Since the size of the minimum cut between $\sigma'$ and $\gamma$ is $sk+\delta$, there are such many edge-disjoint paths from $\sigma'$ to $\gamma$. Then we remove all the edges that are not in the original network $\cN$ from these paths. The resulting paths are the desired ones. 
\end{IEEEproof}

To present our coding scheme, we need more notations. Let $\tbf_{e}\in\F_q^{sk+|\cE|}$ be the extended global encoding vector of link $e$ defined as in Section~\ref{sec:prelim}. The components of $\tbf_e$ can be indexed by the set $[sk]\cup\cE$, that is, 
\[\tbf_e=(\tbf_e(d):d\in[sk]\cup\cE).\]
For an error pattern $\rho\subseteq\cE$ and an extended global encoding vector $\tbf_e$, we define three vectors as follows.
\begin{enumerate}
    \item $\tbf_{e}^\rho\in\F_q^{sk+|\rho|}$ is the  vector obtained from $\tbf_e$ by removing all components $\tbf_e(d)$ where $d\notin [sk]\cup\rho$.
    \item $\bff_{e}^\rho\in\F_q^{sk+|\cE|}$ is the vector obtained from $\tbf_{e}$ by replacing all components $\tbf_e(d)$, where $d\notin [sk]\cup\rho$, with $0$.
    \item $\bff_{e}^{\rho^c}\in\F_q^{sk+|\cE|}$ is the vector obtained from $\tbf_{e}$ by replacing all components $\tbf_e(d)$, where $d\in [sk]\cup\rho$, with $0$.
\end{enumerate}

The following theorem shows the attainability of the Singleton-like bound when $T$ is an identity matrix. 

\begin{thm}\label{thm:identity_case} Let $\cN$ be a directed acyclic network. If the field size $q\geq |R(\delta)|$, then there is a  linear network code $
    \cC$ for $\cN$ such that $d_{\min}(\cC,I,k)=\delta+1$, where $\delta=\min_{C\in\Lambda(\cN)}\set{|C|-k|I_C|}$
\end{thm}
\begin{IEEEproof}
    We extend the network $\cN$ by assigning $k$ imaginary  message channels $\set{d_{(i-1)k+1}, d_{(i-1)k+2},\cdots,d_{ik}}$ to each source node $\sigma_i$ and one imaginary error channel $e'$ to the tail of each edge $e\in\cE$. We denote this new network as $\tilde{\cN}$. For each $\rho\in R(\delta)$, let $\mathcal{P}_\rho$ be a set of $(sk+\delta)$ edge-disjoint paths satisfying the property in Corollary~\ref{cor:edge-disjoint_path}. We denote the set of links on the paths in $\mathcal{P}_\rho$ as $\cE_{\rho}$.

    We define a dynamic set of links $CUT_{\rho}$ for each $\rho\in R(\delta)$, and initialize it as
    \[CUT_\rho=\set{d_i; 1\leq i \leq sk}\cup\set{e';e\in\rho}, \]
    where $e'$ is the imaginary error channel to $e$. For all $e\in\cE$, we initialize  $\tbf_e=\Zero$; for all $d\in \set{d_i; 1\leq i\leq sk}\cup\cE'$, we initial $\tbf_{d}=\bOne_d$, where $\bOne_d$ denotes the binary unit vector with the entry labeled by $d$ being `1'. For a set of vectors $V$, we use $\langle V\rangle$ to denote the linear space that is spanned by the vectors in $V$. For any subset $A\subseteq \set{d_i; 1\leq i\leq sk}\cup\cE\cup\cE'$, we define four vector spaces as follows:
    \begin{align*}
        &\tilde{L}(A)\eqdef \langle\{\tbf_e \mid e\in A\}\rangle,\\
        &\tilde{L}^\rho(A)\eqdef \langle\{\tbf_e^\rho \mid e\in A\}\rangle,\\
        &L^\rho(A)\eqdef \langle\{\bff_e^\rho \mid e\in A\}\rangle,\\
        &L^{\rho^c}(A)\eqdef \langle\{\bff_e^{\rho^c} \mid e\in A\}\rangle.
    \end{align*}
    Note that the initialization above implies that $\tilde{L}^{\rho}(CUT_\rho)=\F_q^{sk+|\rho|}$. 

    Next, we update $\tbf_e$ and $CUT_\rho$ from upstream to downstream until $CUT_\rho\subseteq \In(\gamma)$ for all $\rho \in R(\delta)$. For a link $e\in\cE$, denote $i=\tail(e)$.
    If $e\notin\cup_{\rho\in R(\delta)}\cE_{\rho}$, let $\tbf_e=\textbf{1}_e$, and $CUT_\rho$ remains unchanged. If $e\in\cup_{\rho\in R(\delta)}\cE_{\rho}$, we choose a vector $\tbg_e$ such that 
    \[\tbg_e\in \tilde{L}(\In(i)\cup\{e'\})\backslash\cup_{\set{\rho; e\in \cE_{\rho}}}\big(L^\rho(CUT_{\rho}\backslash\{e_{\rho}\})+L^{\rho^c}(\In(i)\cup\{e'\})\big),\]
    where $e_\rho$ is the previous link of $e$ in $\mathcal{P}_\rho$, and the addition represents the sum of two vector spaces. The existence of such a $\tbg_e$ will be shown later. Next, we choose $\tbf_e$ such that
    \begin{equation*}
        \tbf_e=\begin{cases}
            \tbg_e+\mathbf{1}_e & \text{if $\tbg_e(e)=0$,}\\
            \tbg_e(e)^{-1}\cdot\tbg_e & \text{otherwise.}
        \end{cases}
    \end{equation*}
    For the dynamic set $CUT_\rho$, if $e\in CUT_\rho$, update $CUT_\rho=\{CUT_\rho\backslash\{e_{\rho}\}\}\cup\{e\}$. Otherwise, $CUT_\rho$ remains unchanged. 

    After updating $\tbf_e$ for all $e\in \cE$, $CUT_\rho\subseteq\In(\gamma)$ for every $\rho\in R(\delta)$. 

    To show the existence of $\tbg_e$ is equivalent to show that for $q\geq |R(\delta)|$, 
    \[\bigg|\tilde{L}(\In(i)\cup\{e'\})\backslash\cup_{\set{\rho; e\in \cE_{\rho}}}\big(L^\rho(CUT_{\rho}\backslash\{e_{\rho}\})+L^{\rho^c}(\In(i)\cup\{e'\})\big)\bigg|>0.\]
    Let $\ell=\dim(\tilde{L}(\In(i)\cup\{e'\}))$. For every $\rho$ satisfying $e\in \cE_{\rho}$, we have $e_{\rho}\in \In(i) \cup \set{e'}$. Then $\tbf_{e_\rho}\in\tilde{L}(\In(i)\cup\{e'\})$. However, $\tbf_{e_\rho}\notin L^\rho(CUT_{\rho}\backslash\{e_{\rho}\})+L^{\rho^c}(\In(i)\cup\{e'\})$. This is because that $\tbf_{e_{\rho}}=\bff^{\rho}_{e_\rho}+\bff^{\rho^c}_{e_\rho}$, where $\bff^{\rho}_{e_\rho}\notin L^\rho(CUT_{\rho}\backslash\{e_{\rho}\}), \bff^{\rho}_{e_\rho}\notin L^{\rho^c}(\In(i)\cup\{e'\}),$ and $ \bff^{\rho^c}_{e_\rho}\in L^{\rho^c}(\In(i)\cup\{e'\})$. Therefore,
    \begin{equation}\label{eq:spcdim}
    \dim\parenv*{ \tilde{L}(\In(i)\cup\{e'\}) \cap \parenv*{  L^\rho(CUT_{\rho}\backslash\{e_{\rho}\})+L^{\rho^c}(\In(i)\cup\{e'\})}}\leq \ell-1.
    \end{equation}
    Thus, we have 
    \begin{align}
        & \ \bigg|\tilde{L}(\In(i)\cup\{e'\})\backslash\cup_{\set{\rho; e\in \cE_{\rho}}}\big(L^\rho(CUT_{\rho}\backslash\{e_{\rho}\})+L^{\rho^c}(\In(i)\cup\{e'\})\big)\bigg| \notag \\
        =& \ \bigg|\tilde{L}(\In(i)\cup\{e'\})\bigg|-\bigg|\tilde{L}(\In(i)\cup\{e'\})\cap\parenv*{\cup_{\set{\rho; e\in \cE_{\rho}}}\big(L^\rho(CUT_{\rho}\backslash\{e_{\rho}\})+L^{\rho^c}(\In(i)\cup\{e'\})\big)}\bigg| \label{eq-1} \\
        > & \ q^{\ell}-\sum\limits_{\rho\in R(\delta)}q^{\ell-1} \label{eq-2} \\
        \geq & \   q^{\ell-1}(q-|R(\delta)|)\geq  0. \notag
    \end{align}
    Note that \eqref{eq-1} $\geq$ \eqref{eq-2} due to \eqref{eq:spcdim}. Moreover, if the equality did hold, then necessarily $\abs{R(\delta)}=1$, which is impossible since $\delta < \abs{C}$ for any $C \in \Lambda(\cN)$.

    Finally, we need to show that the encoding coefficients $\tbf_e$'s give rise to a linear network code $\cC$ with $d_{\min}(\cC,I,k)=\delta+1$. We will prove this by showing that during the updating process, $\dim(\tilde{L}^{\rho}(CUT_{\rho}))=sk+\delta$ for all $\rho\in R(\delta)$, which in turn implies that $\Delta(\rho) \cap \Phi=\varnothing$, as finally $CUT_{\rho} \subseteq \In(\gamma)$. 
    
    In the initialization, we have $\dim(\tilde{L}^{\rho}(CUT_{\rho}))=sk+\delta$. Consider  a link $e\in\cE$, assume that all links before $e$ have been updated and $\dim(\tilde{L}^{\rho}(CUT_{\rho}))=sk+\delta$. Recall that
    \[\tbg_e\in \tilde{L}(\In(i)\cup\{e'\})\backslash\cup_{\set{\rho; e\in \cE_{\rho}}}\big(L^\rho(CUT_{\rho}\backslash\{e_{\rho}\})+L^{\rho^c}(\In(i)\cup\{e'\})\big).\]
   It follows that $\tbg^{\rho}_e$ and $\set{\tbf_d^{\rho};d\in CUT_{\rho}\backslash\{e_{\rho}\}}$ are linearly independent for any $\rho$ with $e\in \cE_\rho$. Suppose to the contrary that $\tbg^{\rho}_e$ and $\set{\tbf_d^{\rho};d\in CUT_{\rho}\backslash\{e_{\rho}\}}$ are linearly dependent for some $\rho$. Then $\bg^{\rho}_e\in L^\rho(CUT_{\rho}\backslash\{e_{\rho}\})$. Note that $\bg^{\rho^c}_e\in L^{\rho^c}(\In(i)\cup\{e'\})$ as $\tbg_e\in\tilde{L}(\In(i)\cup\{e'\})$. Thus, $\tbg_e=g^{\rho}_e+\bg^{\rho^c}_e$ is a vector in the sum space $L^\rho(CUT_{\rho}\backslash\{e_{\rho}\})+L^{\rho^c}(\In(i)\cup\{e'\})$, which contradicts to the choice of $\tbg_e$. Now, we show that $\tbf_e^{\rho}$ and $\set{\tbf_d^{\rho};d\in CUT_{\rho}\backslash\{e_{\rho}\}}$ are also linearly independent. 
  \begin{enumerate}
    \item If $\tbg_e(e)\neq0$, since $\tbg^{\rho}_e$ and $\{\tbf_d^{\rho} \mid d\in CUT_{\rho}\backslash\{e_{\rho}\}\}$ are linearly independent   and $\tbf_e=\tbg_e(e)^{-1}\tbg_e$, the statement follows directly. 
    \item If $\tbg_e(e)=0$, we claim that $e\notin \rho$ for any $\rho\in R(\delta)$ such that $e\in E_\rho$. Suppose to the contrary that $e\in\rho$, then $e_\rho=e'$. Therefore, we have $\tbf_{e_\rho}=\mathbf{1}_e$ and $\tbf_d(e)=0$ for $d\in CUT_{\rho}\backslash\{e_{\rho}\}$. Since $\tbg_e(e)=0$ and $\dim(\tilde{L}^{\rho}(CUT_\rho))=sk+\delta$, we have $\tbg_e^{\rho}\in\tilde{L}^{\rho}(CUT_\rho\backslash\{e_\rho\})$. This implies that $\tbg_e$ is a vector in the sum space $L^\rho(CUT_{\rho}\backslash\{e_{\rho}\})+L^{\rho^c}(\In(i)\cup\{e'\})$, which also contradicts to the choice of $\tbg_e$. From the claim, it follows that $\tbf_e^{\rho}=\tbg_e^\rho$, which in turn implies that $\tbf_e^{\rho}$ and $\{\tbf_d^{\rho}\mid d\in CUT_{\rho}\backslash\{e_{\rho}\}\}$ are  linearly independent. 
    \end{enumerate}
    Therefore, after updating $CUT_{\rho}$ by replacing $e_\rho$ with $e$, we still have $\dim(\tilde{L}^{\rho}(CUT_{\rho}))=sk+\delta$.
\end{IEEEproof}

\section{Bounds on Robust Computing Capacity}\label{Sec:capacity}

In this section, we consider the robust computing capacity for 
linear target functions. First we have the following cut-set bound.

\begin{lem}
Let $\cN$ be a directed acyclic network, $f(\bx)=\bx \cdot T$ be a linear function with  $T\in \F_q^{s\times l}$, and $\tau$ be a positive integer. Then
\begin{equation}\label{eq:genbound}
C(\cN, f,\tau)\leq\min\limits_{C\in\Lambda(\cN)}\set*{\frac{|C|-2\tau}{\Rank(T_{I_C})}}.
\end{equation}
\end{lem}

\begin{IEEEproof}
This follows directly from \cite[Theorem~III.1]{WeiXuGe23} and the discussion preceding  \cite[Corollary~II.1]{WeiXuGe23}.
\end{IEEEproof}

When $l\in \set{1,s}$, we have the following result.
\begin{thm}\label{thm:main_result}
Let $\cN$ be an arbitrary directed acyclic network and $\tau$ be a positive integer such that $2\tau < \min_{C \in \Lambda(\cN)} \set{\abs{C}}$. If $q$ is sufficiently large, then we have
\[C(\cN, \bOne,\tau)\geq\min_{C\in \Lambda(\cN)}\set{\abs{C} -2\tau}\]
and 
 \[C(\cN, I,\tau)\geq\min\limits_{C\in\Lambda(\cN)}\set*{\left\lfloor \frac{|C|-2\tau}{\abs{I_C}}\right\rfloor }.\]
\end{thm}
\begin{IEEEproof}
    For $T=\bOne$, let $k=\min_{C\in \Lambda(\cN)}\set{\abs{C} -2\tau}$. Theorem~\ref{thm:Sbndtight} shows that there is a linear network code with 
    \[d_{\min}(\cC,\bOne,k)=\min_{C \in \Lambda(\cN)}\set{ \abs{C}-k+1} =2\tau+1.\] 
    Similarly, for $T=I$, let $k=\min_{C\in\Lambda(\cN)}\set*{\left\lfloor \frac{|C|-2\tau}{\abs{I_C}}\right\rfloor}$. Then Theorem~\ref{thm:identity_case} shows that there is a linear network code with 
    \[d_{\min}(\cC,I,k)=\min_{C \in \Lambda(\cN)} \set{\abs{C}-k\abs{I_C}+1} \geq 2\tau+1.\]
    By Theorem~\ref{thm:minimum_distance}, these codes are resilient to $\tau$ errors.
\end{IEEEproof}

Note that  the above results show that linear network coding can achieve the cut-set bound for $l=1$ and achieve the integral part of the the cut-set bound for $l=s$.

When $1<l< s$, linear network coding scheme for the sum function, along with a technique of time-sharing, can be used to derive a lower bound on the robust computing capacity for  a generic  linear function $\bx \cdot T$. 
\begin{thm}
    Let $\cN$ be an arbitrary directed acyclic network and $\tau$ be a positive integer such that $2\tau<\min_{C\in\Lambda(\cN)}\{|C|\}$. Let $T\in\mathbb{F}_q^{s\times l}$ be a matrix of full column rank. If $q$ is sufficiently large, then
    \begin{equation}\label{eq:genlowerbnd}
    C(\cN, T,\tau)\geq\min\limits_{C\in\Lambda(\cN)}\bigg\{\frac{|C|-2\tau}{l}\bigg\}.
    \end{equation}
\end{thm}
\begin{IEEEproof}
    Let $w'=\min_{C\in\Lambda(\cN)}\{|C|-2\tau\}$. For $1\leq i\leq l$, let $T_i$ denote the $i$-th column of $T$. 
    We construct a coding scheme for $T$ that uses the network $l$ times. 
    In the $i$-th use of $\cN$, the sink node computes $\bx\cdot T_i$. 
    By Theorem~\ref{thm:main_result}, there is a linear coding scheme that can compute the function $f(\bx)$  $w'$ times, while tolerating $\tau$ errors. Therefore, our scheme is able to reliably compute the target function $f(\bx)$ $w'$ times by using the network $l$ times, which establishes  the result.
\end{IEEEproof}

\section{Applications in Distributed 
Computing}\label{Sec:application}
In this section, we explore the applications of linear network codes for robust function computation within the context of distributed computing, with a particular focus on the \emph{gradient coding} problem \cite{TanLeiDimKar17,YeEmm18,RavTamTanDim20,2018Halbawi,2019ozfatura}. 
Consider a data set $\Dataset=\{(\bx_i,y_i)\}_{i=1}^D$ with each tuple $(\bx_i,y_i)\in\F^p\times\F$. Numerous machine learning problems wish to solve problems of the following form:
\[\bbta^*=\arg\min\limits_{\bbta\in\F^p}\sum\limits_{i=1}^D L(\bx_i,y_i;\bbta)+\lambda R(\bbta),\]
where $L(\cdot)$ is a loss function and $R(\cdot)$ is a regularization function. The most commonly used approach to solving this  problem involves gradient-based iterative methods. Let 
\[\grad^{(t)}\eqdef \sum_{i=1}^D \nabla L(\bx_i,y_i;\bbta^{(t)})\in\F^p\] be the gradient of the loss function at the $t^{th}$ step. Then the updates to the model are of the form:
\[\bbta^{(t+1)}=h_{R}(\bbta^{(t)},\grad^{(t)}),\]
where $h_R(\cdot)$ is a gradient-based optimizer which also depends on $R(\cdot)$. As the size of the data set increases, computing the gradient $\grad^{(t)}$ can become a bottleneck. One potential solution is to parallelize the computation by distributing the tasks across multiple workers.

Assume that there  are $n$ worker nodes, denoted by $W_1,W_2,\cdots,W_n$, and the data set $\Dataset$ is partitioned into $K$ data subsets, denoted by $\Dataset_1,\Dataset_2,\cdots,\Dataset_K$.  The \emph{partial gradient vector} $\grad_i^{(t)}$ is defined as 
\[\grad_i^{(t)}\eqdef  \sum_{(\bx,y)\in \Dataset_i} \nabla L(\bx,y;\bbta^{(t)}).\] 
Then
\[\grad^{(t)}=\grad_1^{(t)}+\grad_2^{(t)}+\cdots \grad_k^{(t)}.\]

The master node initially assigns the data subsets $\Dataset_1,\Dataset_2,\cdots,\Dataset_K$ to the worker nodes. Let $\Asnset_i$ denote the set of indices corresponding to the data subsets stored by worker node $W_i$. Each worker $W_i$ computes the partial gradients $\set*{\grad_j^{(t)};j\in \Asnset_i}$ based on its assigned data subsets and then transmits a coded message $f_i(\grad_j^{(t)}: j\in \Asnset_i) \in \F^{p/m}$ to  the master node, where $f_i$ is a linear function that encodes the partial gradients in $\Asnset_i$, and $m$ is referred to as  \emph{communication reduction factor}. Due to stragglers—worker nodes slowed down by unpredictable factors such as network latency—the master node may not receive all the coded messages, but rather  $n-\tau$ of them. It must then decode the sum of partial gradients $\grad^{(t)}$ from these received messages.
The primary problem is designing a gradient coding scheme that includes data assignment and message encoding/decoding to increase straggler tolerance $\tau$ while minimizing communication cost and computation cost. The communication cost can be parameterized by $1/m$  while the computation cost can be parameterized by the number of worker nodes that each data subset is assigned. Since the encoding functions $f_i$'s are time invariant, we omit the superscript $(t)$ in the rest of this paper for simplicity of notation.

The authors in \cite{TanLeiDimKar17} characterized the trade-off between the straggler tolerance and the computation cost when the communication reduction factor $m=1$. A gradient coding scheme which can achieve this trade-off was also proposed. This scheme consists of a \emph{cyclic} data assignment, where each worker node $W_i$ is assigned with $\Dataset_i, \Dataset_{i+1},\ldots,\Dataset_{i+w}$ for some fixed $w$, and a random code construction. Subsequently, a deterministic code construction based on cyclic MDS codes was proposed in \cite{RavTamTanDim20} to replace the random code construction in \cite{TanLeiDimKar17}. For general $m\geq 1$, the authors in \cite{YeEmm18} characterized the optimal trade-off between straggler tolerance, computation cost and communication cost. Among others, they proved the following converse bound. 
\begin{lem}[{\cite[Appendix A]{YeEmm18axv}}]\label{lm:convbnd}
In a gradient coding scheme with $n$ worker nodes, $K$ data subsets with communication reduction factor $m$ and straggler tolerance $\tau_s$, every data subsets must be assigned to at least $\tau_s+m$ worker nodes. 
\end{lem}
A gradient coding scheme that achieves the converse bound was also proposed, where each data subset is assigned to exactly $\tau_s+m$ worker nodes, and each worker node is assigned $\tau_s+m$ data subsets.  

In the literature on gradient coding, it is typically assumed that the system is homogeneous, meaning all worker nodes have the same storage capacity and computation speed. Consequently, in all the aforementioned works, each worker node is assigned the same number of data subsets. 
In this section, we consider a \emph{heterogeneous} scenario where the worker nodes have varying storage capacities and computation speeds. Intuitively, worker nodes with lower storage capacity and computation speed should be assigned less data to avoid slowing down the overall computation time.

In the following, we first show that for an arbitrary data assignment, network coding can be used to design the encoding functions $f_i$'s for the worker nodes, enabling the gradient coding scheme to achieve the converse bound stated in Lemma~\ref{lm:convbnd}. Then, we show how to design the data assignment to accommodate the heterogeneous scenario.

For a given data assignment $\mathscr{Z}=\set{\Asnset_i;1\leq i \leq n}$, we can construct a three-layer network $\cN(\mathscr{Z})$ as follows. The nodes in the first layer are labeled by the data subsets $\Dataset_1,\Dataset_2,\ldots,\Dataset_{K}$, the nodes in the middle layer are labeled by the worker nodes $W_1,W_2,\ldots,W_n$, and the sink node corresponds to the master node. There is a link from a node  labeled by $\Dataset_i$ to a node labeled by $W_j$ if and only if $\Dataset_i$ is assigned to $W_j$. Additionally, there is a link from each node labeled by $W_j$ to the sink node. We treat each partial gradient $\grad_i$ as a message generated at the node $\Dataset_i$. Since any worker node storing $\Dataset_i$ can compute coded message of $\grad_i$, the problem of designing a gradient coding scheme can be reduced to designing a network coding scheme that enables the sink node to compute the sum $\grad=\sum_{i=1}^k \grad_t$ even if there are $\tau_s$ outages in the incoming links to the sink node.

\begin{prop}\label{pro:NFCC2GC}
Let $\tau_s$ and $m$ be positive integers.
For the three-layer network $\cN(\mathscr{Z})$, suppose that there is a linear network code $\cC$ which can compute the sum function with $d_{\min}(\cC,\bOne,m)\geq \tau_s+1$. Then there is a gradient coding scheme, incorporating $\mathscr{Z}$ as the data assignment, with straggler tolerance $\tau_s$ and communication reduction factor $m$.
\end{prop}

\begin{IEEEproof} For each partial gradient $\grad_j$, we write it as\footnote{We treat the partial gradient $\grad_j$ as a row vector.} $\grad_j = (\grad_j(1),\grad_j(2),\ldots,\grad_j(p/m))$, where each $\grad_j(\ell)\in \F^{m}$.
Let $F$ be the global encoding matrix of $\cC$. Since  there are $K$ source nodes and $n$ incoming links to the sink node, then $F\in \F^{(mK)\times n}$. For $1\leq i\leq n$, let $\bff_i$ be the column of $F$ that corresponds to the link from the node labeled by $W_i$ to the sink node. 
Noting that the nonzero entries of $\bff_i$ correspond to the source nodes labeled by $\Dataset_j$ such that $j \in \Asnset_i$, we define the gradient encoding function for the worker $W_i$ as 
\[f_i(\grad_j:j\in \Asnset_i) \eqdef \parenv*{ \parenv*{\grad_1(\ell),\grad_2(\ell),\ldots,\grad_K(\ell)}\cdot \bff_i :1\leq \ell \leq p/m} \in \F^{p/m}.\]

Since $d_{\min}(\cC,\bOne,m)\geq \tau_s+1$, according to Remark~\ref{rmk.erasure},  even if there are $\tau_s$ outages in the incoming links of the master node, it still can decode the sum $\sum_{i=1}^K \grad_i(\ell)$, for all $1\leq \ell \leq p/m$, and so, $\sum_{i=1}^K \grad_i$.
\end{IEEEproof}

\begin{thm} Let $\tau_s$ and $m$ be positive integers.  
Let $\mathscr{Z}$ be a data assignment with $n$ worker nodes and $K$ data subsets such that every data subset is assigned to at least $\tau_s+m$ worker nodes. Then there is a gradient coding scheme incorporating $\mathscr{Z}$ with communication reduction factor $m$ and straggler tolerance $\tau_s$.
\end{thm}

\begin{IEEEproof} Since each data subset is assigned to at least $\tau_s+m$ worker nodes, the minimum-degree of the source nodes in $\cN(\mathscr{Z})$ is at least $\tau_s+m$. By Theorem~\ref{thm:3layernetwork}, there is a linear network code $\cC$ with $d_{\min}(\cC,\bOne,m) =\tau_s+1$ as the field $\F$ is sufficiently larger. The conclusion then follows from Proposition~\ref{pro:NFCC2GC}.
\end{IEEEproof}

In order to design a data assignment  accommodating the heterogeneous scenario, we use the approach in \cite{WooCheJi21}. For a collection of worker nodes $W_1,W_2,\ldots,W_n$, we use a vector $\br=(r_1,r_2,\ldots,r_n)\in \Q^n$ and a vector $\bs=(s_1,s_2,\ldots,s_n) \in \Q^n$ to represent the storage capacity and computation speed, respectively, where $r_i$ is the fraction of the data that is assigned to $W_i$ and  $s_i$ is the fraction of data that $W_i$ can compute per unit time.

For a collection of data subsets $\mathscr{D}=\set{\Dataset_1,\Dataset_2,\ldots, \Dataset_K}$ and a data assignment $\mathscr{Z}=\set{\Asnset_1,\Asnset_2,\ldots,\Asnset_n}$, the vector $\bmu=(\mu_1,\mu_2,\ldots,\mu_u) \in \Q^n$, where  
\[\mu_i\eqdef \frac{\sum_{j\in \Asnset_i} \abs{\Dataset_j}}{\sum_{j=1}^K \abs{\Dataset_j}},\]
is the computation load vector. Our goal is to minimize the overall computation time  
\[c(\mathscr{D},\mathscr{Z})\eqdef \max_{1\leq i\leq n}\frac{\mu_i}{s_i},\]
while ensuring that each data subset is assigned to at least $\tau_s+m$ worker nodes.

This problem can be formulated as the following optimization problem:
\begin{align}
\mathop{\textrm{minimize }}\limits_{\mathscr{D},\mathscr{Z}} \quad & c(\mathscr{D},\mathscr{Z}) \\
\textrm{subject to } \quad &  \mu_i \leq r_i \textrm{ for all } 1\leq i\leq n,\\
& \abs{\set{i ; j \in \Asnset_i}} \geq \tau_s+m \textrm{ for all } 1\leq j \leq K. \label{eq:condrep}
\end{align}
To solve this problem, we adopt the approach in \cite{WooCheJi21} and decompose it into two sub-problems. The first one is the following relaxed convex optimization problem to find
the optimal computation load vector $\bmu^*$:
\begin{align*}
\mathop{\textrm{minimize }}\limits_{\bmu} \quad & \max_{1\leq i\leq n} \frac{\mu_i}{s_i} \\
\textrm{subject to } \quad &  \mu_i \leq r_i \textrm{ for all } 1\leq i\leq n,\\
& \sum_{i=1}^n \mu_i  \geq \tau_s+m.
\end{align*} The solution to this problem can be found in \cite[Theorem~1]{WooCheJi21}. The second problem is to find data assignment scheme $\mathscr{Z}$, as well as data partition $\mathscr{D}$, with computation load vector $\bmu^*$ such that \eqref{eq:condrep} holds. This is solved in \cite[Section V]{WooCheJi21}, using the fact that $\sum_{i=1}^n \mu^*_i\geq \tau_s+m$.

Recently, the  gradient coding problem was extended in \cite{WanSunJiCai22} to compute a \emph{linearly separable function} $f$, which can be written as 
\[f(\Dataset_1,\Dataset_2,\ldots,\Dataset_K)=g(f_1(\Dataset_1), f_2(\Dataset_2), \ldots, f_K(\Dataset_K)),\]
where $g$ is a linear map defined by $l$ linear combinations of $f_i(\Dataset_i)$'s.

Given the straggler tolerance $\tau_s$,  the authors in \cite{WanSunJiCai22} examined the specific case where the computation cost is
minimum, and proposed novel schemes along with converse bounds for the optimal communication cost. The proposed scheme is optimal under the constraint of cyclic data assignment. 
However, it is unknown whether this  scheme remains optimal if this constraint is removed. Therefore, it is of particular interest to investigate coding schemes for other data assignments.

Using the same reasoning  as in Proposition~\ref{pro:NFCC2GC}, this problem can  be translated into a robust function computation  problem in a three-layer network with the target function $f(\bx)=\bx \cdot T$, where $T\in \F^{s\times l}$.  However, in a generic three-layer network, this  problem remains open  when $2\leq l \leq s-1$.

In this section, we have treated stragglers as communication outages and used linear network coding with a distance of at least $\tau_s+1$ to mitigate their impact. It is worth noting that this approach can also be applied to defend against Byzantine attacks, where some worker nodes send misleading or incorrect messages to the master node, causing computation errors. In this case, we treat the incorrect messages as erroneous links and assume there are at most $\tau_m$ malicious nodes. Theorem~\ref{thm:minimum_distance} guarantees  that a linear network code with a distance of at least $2\tau_m+1$ can effectively counter such attacks. Similar to Proposition~\ref{pro:NFCC2GC}, we have the following result, the proof of which is analogous and thus omitted here.

\begin{prop}
Let $\tau_b$ and $m$ be positive integers.
For the three-layer network $\cN(\mathscr{Z})$, suppose there exists a linear network code $\cC$ that  computes the target function $f(\bx)=\bx \cdot T$ with $d_{\min}(\cC,T,m)\geq 2\tau_b+1$. Then there is a coding scheme, incorporating $\mathscr{Z}$ as the data assignment, which has communication reduction factor $m$ and can tolerate up to $\tau_b$ malicious nodes.
\end{prop}

\bibliographystyle{IEEEtranS}
\bibliography{reference}

\end{document}